\documentclass[aps,prd,onecolumn,groupedaddress,amssymb,showpacs,nofootinbib,epsfig,pdffig]{revtex4}
\usepackage{graphicx}
\usepackage{bm}
\usepackage{dcolumn}
\usepackage{amsmath}
\usepackage{amssymb}
\usepackage{epsfig}
\usepackage{color}

\def \lleq {\lower0.9ex\hbox{ $\buildrel < \over \sim$} ~}
\def \ggeq {\lower0.9ex\hbox{ $\buildrel > \over \sim$} ~}

\newcommand{\rd}{{\rm d}}

\def \omm  {\Omega_{0 {\rm m}}}

\def \beq  {\begin{equation}}
\def \eeq  {\end{equation}}
\def \ber  {\begin{eqnarray}}
\def \eer  {\end{eqnarray}}

\def\mn{{Mon.\@ Not.\@ Roy.\@ Ast.\@ Soc.\ }}

\def\plb {{Phys.\@ Lett.\@ B\ }}
\def \jetpl {JETP Lett.\ }

\newcommand{\be}{\begin{equation}}
\newcommand{\ee}{\end{equation}}
\newcommand{\ba}{\begin{eqnarray}}
\newcommand{\ea}{\end{eqnarray}}
\newcommand{\bea}{\begin{eqnarray*}}
\newcommand{\eea}{\end{eqnarray*}}

\begin{document}

\title{Cosmological dynamics of non-minimally coupled scalar field system and its \\late time cosmic relevance
 }

\author{M. Sami}
\affiliation{Kobayashi-Maskawa Institute for the Origin of Particles
and the Universe, \\Nagoya University, Nagoya 464-8602, Japan}
\affiliation{Centre for Theoretical Physics, Jamia Millia Islamia,
New Delhi-110025, India}
\author{  M. Shahalam }
\affiliation{ Centre for Theoretical Physics, Jamia Millia Islamia,
New Delhi-110025, India}

\author{M. Skugoreva}
\affiliation{Institute of Gravitation and Cosmology, Peoples
Friendship\\ University of Russia, Moscow 117198, Russia}

\author{A. Toporensky}

 \affiliation{ Sternberg Astronomical Institute,
Moscow 119992, Russia}

\begin{abstract}
We investigate the cosmological dynamics of non-minimally coupled
scalar field system described by $F(\phi)R$ coupling with
$F(\phi)=\left( 1-\xi\phi^N\right)R$($N\ge2$) and the field
potential, $V(\phi)=V_0\phi^n$. We use a generic set of dynamical
variables to bring out new asymptotic regimes of the underlying
dynamics. However, our dynamical variables miss the most important
fixed point$-$ the de Sitter solution. We make use of the original
form of system of equations to investigate the issues related to
this important solution. In particular, we show that the de-Sitter
solution which is a dynamical attractor of the system lies in the
region of negative effective gravitational constant $G_N$ thereby
leading to a ghost dominated universe in future and a transient
quintessence(phantom) phase  with $G_N>0 $ around the present
epoch\footnote{However, as demonstrated by Starobinsky in 1981, the
ghost dominated universe, if exists, can not be accessed from the
Universe we live in, we shall say more about this important result
in the last section.}. We also carry out comparison of the model
with other competing models of dark energy such as galileon modified
gravity and others.

\end{abstract}
\maketitle

\section{Introduction}
Theories with a scalar field non-minimally coupled to gravity dubbed
scalar tensor theories have been studied for decades. The first
well-known example of non-minimal coupling {\it a la} Brans-Dicke
theory, was proposed in 1961 with an aim to match Mach principle
with General Relativity \cite{BD}. In this theory the gravitational
constant is replaced by a scalar field $\phi$ entering into the
action in a specific combination with Riemanian curvature as $\phi^2
R$.

Followed by the Brans-Dicke proposal, other forms of scalar-tensor
action were investigated, a well known example of a non-minimally
coupled system is provided by  $F(\phi)R$ coupling with $F=1-\xi
\phi^2$. The cosmological dynamics of such a theory is rather rich
and deserves attention. For a recent development in this direction,
it is worth nothing that non-minimally coupled Higgs field due to a
large coupling $\xi$ might give rise to a successful inflation
\cite{BS} which is otherwise impossible.

 The non-minimally coupled scalar field system due to novel features are of great interest
 to dark energy model building\cite{PR1,PR2,review1,vpaddy,review2,review3,review3C,review3d,review4}.
 For instance, non-minimal coupling might allow
 phantom crossing and may give rise to cosmological scaling solutions
 of interest to models of dark energy. Phantom scaling solutions are
 generic features of a non-minimally coupled system with $F=1-\xi
 \phi^2$\cite{Polarski}.

In recent years, methods of dynamical system theory have been
extensively used in cosmology for obtaining  a general picture of
dynamics for many cosmological models including those with a scalar
field and modified gravity. The advantage of this method is in
having some kind of "machinery " for deriving asymptotic solution
using a simple programmed algorithm. This requires introduction of
new set of variables in which the initial system can be rewritten as
a system of first-order equations. On the other hand, there is a
danger of losing some important solution as well as inability for
this scheme to find transient regimes which can also be important.
Nevertheless, a classification of stable asymptotic regimes, given
by this method may be useful for understanding the underlying
dynamics. The success and the limitation of the framework of
dynamical systems applied
 to $f(R)$ theory can be found in Refs.\cite{Tsujikawa} and \cite{Dunsby}.

 We shall restrict our discussion to a polynomial functions $F$ of the form
$F(\phi)=1-\xi \phi^N$ ($N\geq2$) and power-law potentials for the
scalar field, $V(\phi)=V_0\phi^n$ giving rise to generalization of
models considered earlier\cite{Polarski,Dunsby,Dunsby1,mark}(see
also Ref.\cite{sergei} on the related theme). The set of variables
that we use can help in bringing out some generic features of the
underlying dynamics and new asymptotic regimes missed in earlier
studies. We should, however, note that the set of variables used in
the present paper is not useful for the study of approximate
Einstein regime in the system under consideration. For detailed
description of  this regime, other methods are required
\cite{Kuusk}. Secondly, our variables miss the existence of de
Sitter solution  and in order to investigate its existence and
stability, we need to go back to initial variables to perform the
analysis.

 In this paper, we  investigate cosmological dynamics of
 non-minimally coupled scalar field system with  specific functional forms of coupling
 and field potential
 using a convenient set
 of dynamical variables. We shall focus on the asymptotic regimes of the solutions of
 interest and reveal the important features associated with the de
 Sitter solution.
 We also carry out comparison of the model
 with other competing models of dark energy, in particular, the galileon system which is
 generically non-minimally coupled.

\section{Equations of motion}
\label{sec1}

Let us consider the scalar field system with non-minimal coupling in
the form,

 \be
 \label{eq:Lagrangian}
S=\frac{1}{2}\int{\sqrt{-g}d^4x\Big{[} m_{Pl}^2
R-(g^{\mu\nu}\phi_{\mu}\phi_{\nu}+ \xi R
B(\phi)+2V(\phi))\Big{]}}+S_M, \ee where $m_{Pl}^2=1/{8\pi
G}=1/{\kappa}$, and  $\xi$ is the dimensionless parameter and $S_M$
designates matter action.

In a homogenous isotropic Friedmann-Robertson-Walker universe  with
spatially flat metric, \be \label{eq:metric} ds^2
=-dt^2+a^2(t){dl}^2, \ee

the equations of  motion which are obtained by varying the action
(\ref{eq:Lagrangian}) have the form,

\be \label{eq:Friedphi} H^2=\frac{\kappa}{3}\left(\frac{1}{2}{\dot
{\phi}}^{2}+V(\phi)+3\xi(H \dot {\phi} B'(\phi)+H^{2}B(\phi))+\rho
\right), \ee
\begin{eqnarray}
\label{eq:Friedphi2} R=\kappa \left(-{\dot {\phi}}^{2} +4
V(\phi)+3\xi(3 H \dot{\phi}  B'(\phi)+\frac{R}{3}  B(\phi)  + {\dot
{\phi}}^{2}B''(\phi)+\ddot {\phi} B'(\phi))+\rho(1-3\omega) \right),
\end{eqnarray}
\begin{eqnarray}
\label{eq:KGphi}
&&\ddot {\phi}+3 H \dot {\phi}+\frac{1}{2}\xi R B'(\phi)+V'(\phi)=0.
\end{eqnarray}
where $\rho$ and $p$ are the energy density and pressure of the
ordinary matter with $p=\omega \rho$.

From the standard form of equations
\begin{eqnarray*}
R_{ij}-\frac{1}{2}R g_{ij}&=&8\pi G_N(T_{ij,\phi}+T_{ij,m})=\kappa
T^{eff}_{ij} ,
\end{eqnarray*}
we read off the expression for $G_N$ as the effective Newtonian
gravitational constant
\begin{equation}
G_N=\frac{\kappa}{8\pi(1-\kappa \xi B(\phi))},
\end{equation}
and we shall use this definition of $G_N$ hereafter. Ricci Scalar in
chosen metric is given by, $R=6(2H^2+ \dot {H})$. For convenience,
we shall use the system of unites with $\kappa=6$ .

Dividing (\ref{eq:Friedphi}), (\ref{eq:Friedphi2}) ,(\ref{eq:KGphi})
by $H^2(1-6\xi B(\phi))$ and  multiplying (\ref{eq:KGphi}) by $\xi
B'(\phi)$, we obtain,
\begin{eqnarray}
\label{eq:newFried1} 1=\frac{{\dot {\phi}}^{2} }{H^2(1-6\xi
B(\phi))}+\frac{2 V(\phi)}{H^2(1-6\xi B(\phi))} +\frac{6\xi \dot
{\phi} B'(\phi)}{H(1-6\xi B(\phi))}+\frac{2\rho}{H^2(1-6\xi
B(\phi))},
\end{eqnarray}
\begin{eqnarray}
\label{eq:newFried2} \frac{R}{H^2}&=&-\frac{6 {\dot {\phi}}
^{2}}{H^2(1-6 \xi B(\phi))}+\frac{24 V(\phi)}{H^2(1-6 \xi B(\phi))}
+\frac{54 \xi \dot {\phi} B'(\phi)}{H(1-6\xi B(\phi))}+\frac{18 \xi
{\dot{\phi}} ^2 B''(\phi)}{H^2(1-6 \xi B(\phi))}\nonumber\\
&+&\frac{18 \xi \ddot {\phi}  B'(\phi)}{H^2(1-6 \xi B(\phi))}
 +\frac{6\rho(1-3\omega)}{H^2(1-6 \xi B(\phi))},
\end{eqnarray}
\begin{eqnarray}
\label{eq:newKG} 0=\frac{\xi \ddot {\phi} B'(\phi)}{H^2(1-6 \xi
B(\phi))}+\frac{3 \xi \dot {\phi} B'(\phi)}{H(1-6 \xi B(\phi))}
+\frac{R}{H^2}\frac{\xi^{2}{B'}^2(\phi)}{2(1-6 \xi B(\phi))}
+\frac{V'(\phi)\xi B'(\phi)}{H^2(1-6 \xi B(\phi))}.
\end{eqnarray}
In the discussion to follow, we shall use the dimensionless
variables,
\begin{eqnarray}
\label{eq:omega} x &=& \frac{{\dot {\phi}}^{2}}{H^2(1-6 \xi
B(\phi))}, ~~y = \frac{2 V(\phi)}{H^2(1-6 \xi B(\phi))},~~
z = \frac{6\xi \dot {\phi} B'(\phi)}{H(1-6 \xi B(\phi))},\nonumber\\
\Omega &=& \frac{2 \rho}{H^2(1-6 \xi B(\phi))},
\end{eqnarray}
and the dimensionless parameters that depend on the specific form of
functions $B(\phi)$, $V(\phi)$,
\begin{eqnarray}
A=\frac{B'(\phi)\phi}{(1-6 \xi B(\phi))},~~
b=\frac{B''(\phi)\phi}{B'(\phi)},~~ c=\frac{V'(\phi)\phi}{V(\phi)}.
\end{eqnarray}
Here $'$ denotes derivative with respect to $\phi$.


Using the new variables and parameters, we  can now cast the
equations of motion (\ref{eq:newFried1}), (\ref{eq:newFried2}),
(\ref{eq:newKG}) in the autonomous form,


\begin{eqnarray}
\label{eq:autonomous4}
\frac{\dot{x}}{H}&=&\frac{dx}{d\ln a} = x' = 12X\frac{x}{z}-2x(\frac{Y}{6}-2)+xz,\nonumber\\
\frac{\dot{y}}{H}&=&\frac{dy}{d\ln a} = y' = \frac{yz}{6\xi}\frac{c}{A}-2y(\frac{Y}{6}-2)+yz,\nonumber\\
\frac{\dot{z}}{H}&=&\frac{dz}{d\ln a} = z' = 6X+\frac{z^2}{6\xi}\frac{b}{A}-z(\frac{Y}{6}-2)+z^2,\nonumber\\
\frac{\dot{A}}{H}&=&\frac{dA}{d\ln a} = A' = \frac{z}{6\xi}(b+1)+Az,\nonumber\\
\frac{\dot{\Omega}}{H}&=&\frac{d\Omega}{d\ln a} = {\Omega}^{'} =
\Omega(-3-3\omega-2(\frac{Y}{6}-2)+z),
\end{eqnarray}
where we  have written equations for variables and parameters $x$,
$y$, $z$, $A$, $\Omega$ and dropped  $\dot{b}$ $\&$ $\dot{c}$ as we
would be interested in the cases of $b,c=const$ (see Sec.
\ref{stationary}). While writing equation for ${{\Omega}}$, we used
the continuity equation, $\dot{\rho}+3H\rho(1+\omega)=0$.
Secondly, the higher derivative terms  involving $\dot H$  and
$\ddot {\phi}$ in the equations of motion,
\begin{eqnarray}
\label{eq:xyr} X\equiv\frac{\xi \ddot {\phi} B'(\phi)}{H^2(1-6 \xi
B(\phi))},~~~ Y\equiv\frac{R}{H^2},
\end{eqnarray}
and $\Omega$ get expressed through  autonomous variables $x,y,z$ and
are given by,

\begin{eqnarray}
\label{eq:capxy}
\Omega&=&1-x-y-z,\nonumber\\
X(x,y,z)&=&-\frac{z}{2}-\frac{z^2}{18(4x+z^2)}\left(-6x+12y+\frac{z^2
b}{2\xi A}+\frac{yc}{\xi A}\right.
 + 3 (1-x-y-z)(1-3\omega) \Big{)},\nonumber\\
Y(x,y,z)&=&\frac{4x}{4x+z^2}\left(-6x+12y+\frac{z^2}{4\xi A}
\left(2b-\frac{yc}{x}\right)\right. + 3 (1-x-y-z)(1-3\omega)\Big{)}.
\end{eqnarray}
Substituting these expressions in the system (\ref{eq:autonomous4}),
we finally obtain,
\begin{eqnarray}
\label{eq:prime1}
x'&=& 12x\left(-\frac{1}{2}-\frac{z}{18(4x+z^2)}(-6x+12y+\frac{z^2 b}{2\xi A}+\frac{yc}{\xi A}\right. \nonumber \\
&&+ 3(1-x-y-z)(1-3\omega)) \Big{)}-2x\left(\frac{2x}{3(4x+z^2)}\right.\nonumber \\
&& (-6x+12y  + \frac{z^2}{4\xi A} (2b-\frac{yc}{x})+3(1-x-y-z)\nonumber \\
&&(1-3\omega))-2\Big{)}+xz,\nonumber \\
y'&=&\frac{yz}{6\xi}\frac{c}{A}-2y\left(\frac{2x}{3(4x+z^2)} (-6x+12y+\frac{z^2}{4\xi A} (2b-\frac{yc}{x} )\right.\nonumber\\
&& + 3(1-x-y-z)(1-3\omega) )-2\Big{)}+yz,\nonumber \\
z'&=&6\left(-\frac{z}{2}-\frac{z^2}{18(4x+z^2)}(-6x+12y+\frac{z^2 b}{2\xi A}+\frac{yc}{\xi A}\right.\nonumber \\
&&+3(1-x-y-z)(1-3\omega)) \Big{)}-z\left(\frac{2x}{3(4x+z^2)}\right.\nonumber \\
&&(-6x+12y +\frac{z^2}{4\xi A}(2b-\frac{yc}{x})+3(1-x-y-z)\nonumber \\
&&(1-3\omega))-2\Big{)}+\frac{z^2}{6\xi}\frac{b}{A}+z^2,\nonumber \\
A'&=&\frac{z}{6\xi}(b+1)+Az.
\end{eqnarray}
 Let us note that $z$ appearing in the denominator in
(\ref{eq:autonomous4}) cancels upside down after the substitution of
(\ref{eq:capxy}) into (\ref{eq:autonomous4}) thereby telling us that
$z=0$ is a regular point of the dynamical system.

In what follows, we shall investigate the autonomous system
(\ref{eq:autonomous4}) for fixed points. We would specially be
interested in stable solutions of interest to late time cosmic
acceleration. For simplicity, we shall assume specific functional
forms for functions $B(\phi)$ and $V(\phi)$.

\section{Stationary points and their stability : $B(\phi) = {\phi}^N$, $V(\phi) = V_0{\phi}^n$}
\label{stationary}
  In this special case,
  $b={B''(\phi)\phi}/{B'(\phi)}=N-1$ and
  $c={V'(\phi)\phi}/{V(\phi)}=n$. Our autonomous system was
  written keeping this simple case in mind.
   We shall find stationary points equating to zero the left-hand sides of the system (\ref{eq:autonomous4}).
   Their stability will be established using the sign of the corresponding eigenvalues which we shall
   obtain numerically. We begin our discussion  from the case, $N\neq2$ ($b\neq1$).
 In the case of $N=2$, a simple additional relation exists between
$x$ and  $z$,
 this case will be  taken up  separately in the Appendix. In the general case the form of this relation is more
involved algebraically and its substitution into our system leads to
 cumbersome equations. We, therefore, prefer not to make use of it, we would rather check
resulting solutions for consistency.


\subsection{The case of $b\neq1(N\neq2)$.}

After solving the system of algebraic equations obtained after
equating the left hand sides of
(\ref{eq:autonomous4}) to zero, we find the following stationary points:\\
\\\textbf{1. Vacuum stationary line:}~~~~$x=1, y=0, z=0,
A\in(-\infty, +\infty),  \Omega=0$\\

This is a vacuum solution for which the corresponding eigenvalues
are given by,
\begin{eqnarray}
{\lambda}_1=3-3\omega,~~ {\lambda}_2=6,~~ {\lambda}_3=0,~~
{\lambda}_4=0.
\end{eqnarray}
As one of these eigenvalues is positive, the stationary line is
unstable for any value of $\xi$ and $\omega$.

The time dependence of scale factor can be found using the second
relation in (\ref{eq:xyr}),
\begin{eqnarray}
Y_{stat}&=&\frac{R}{H^2}=6\left(2+\frac{\dot{H}}{H^2}\right)\nonumber
\end{eqnarray}
Since $Y_{stat}$ is constant, the above equation can easily be
integrated to obtain the expression for $a(t)$
\begin{eqnarray}
a(t)&=&a_0{|t-t_0|}^{\frac{1}{2-\frac{Y_{stat}}{6}}},\nonumber
\end{eqnarray}
where $t_0$ is integration constant. Since $Y_{stat}=-6$ for the
stationary point under consideration, we finally have
\begin{eqnarray}
a(t)&=&a_0{|t-t_0|}^{\frac{1}{3}}.
\end{eqnarray}
{ We note that at all stationary points either $t\rightarrow t_0$ or
$t\rightarrow\infty$ which means that starting from anywhere in the
phase space, the fixed points are approached in the said limits}.
And all functions of time ($a(t)$, $\phi(t)$, $H(t)$, $\rho(t)$)
either go to zero or infinity or become constants in this limit.
For instance, the stationary line coordinate $A$ has the following limiting values\\
$$
\begin{cases}
0, \text{if $\phi(t)\rightarrow 0,$}\\
-\frac{N}{6\xi}, \text{if $\phi(t)\rightarrow\infty$}\\
\frac{{N \phi_0}^N}{1-6\xi{\phi_0}^N}, \text{if $\phi=\phi_0=const$}.
\end{cases}
$$
1).  In case  of $\phi(t)\rightarrow 0$ and $A\rightarrow 0$, we
find $\phi(t)$ using the coordinate $x$ of this stationary line. As
$(1-6\xi\phi^N)\rightarrow 1$ for $\phi(t)\rightarrow 0$, we have
\begin{equation}
\begin{array}{l}
x_{stat}\rightarrow\frac{{\dot{\phi}}^2}{H^2},\\
\pm\dot{\phi}=\sqrt{x_{stat}}H=\pm\sqrt{x_{stat}}\frac{\dot{a}}{a},\\
\phi(t)=\pm\frac{\sqrt{x_{stat}}}{2-\frac{Y_{stat}}{6}}\ln
|\frac{t-t_0}{t'-t'_0}|,
\end{array}
\end{equation}
where
\begin{eqnarray}
x_{stat}>0, \phi'_0=const,~~
t'-t'_0=\frac{{a'_0}^{2-\frac{Y_{stat}}{6}}}{{a_0}^{2-\frac{Y_{stat}}{6}}
e^{\phi'_0\frac{2-\frac{Y_{stat}}{6}}{\sqrt{x_{stat}}}}}=const.\nonumber
\end{eqnarray}
    For the stationary line under consideration, $x_{stat}=1$, $Y_{stat}=-6$
    which gives
\begin{equation}
\begin{array}{l}
\phi(t)=\pm\frac{1}{3}\ln |\frac{t-t_0}{t'-t'_0}|,
\end{array}
\end{equation}

    Thus $\phi(t)\rightarrow\pm\infty$ for $t\rightarrow t_0$ or $t\rightarrow\infty$.
    The obtained behavior of $\phi(t)$
    contradicts  the initial assumption that $\phi(t)\rightarrow 0$
    and this case should therefore be discarded from the
    discussion.\\
\\

2). In case $\phi(t)\rightarrow\infty$ and
$A\rightarrow-\frac{N}{6\xi}$,  we can find $\phi(t)$ using the
following combinations of stationary point coordinates
\begin{equation}
\begin{array}{l}
\label{eq:point2in1}
\frac{x_{stat}}{A_{stat}}=\frac{{\dot{\phi}}^2}{H^2 N{\phi}^N}=\frac{{\dot{\phi}}^2 a^2}{N{\phi}^N{\dot{a}}^2}=\gamma_{stat},\\
\\\frac{{\dot{\phi}}^2}{{\phi}^N}=N \gamma_{stat}{\left( \frac{\dot{a}}{a}\right) }^2,\\
\\\text{where either $\gamma_{stat}>0$, ${\phi}^N>0$ or $\gamma_{stat}<0$, ${\phi}^N<0$}\\
\\\pm\frac{{|\phi|}^{1-\frac{N}{2}}}{(1-\frac{N}{2})}-{\phi'}_0=\sqrt{N |\gamma_{stat}|}\ln|\frac{a}{a'_0}|,
\\{|\phi(t)|}^{\frac{2-N}{2}}=\pm \frac{3(2-N)\sqrt{N |\gamma_{stat}|}}{12-Y_{stat}}\ln |\frac{t-t_0}{t'-t'_0}|,
\end{array}
\end{equation}
where
\begin{equation}
 t'-t'_0=\frac{{a'_0}^{2-\frac{Y_{stat}}{6}}}{{a_0}^{2-\frac{Y_{stat}}{6}}
e^{\phi'_0\frac{2-\frac{Y_{stat}}{6}}{\sqrt{N
|\gamma_{stat}|}}}}=const,~~\gamma_{stat}=const,~~ \phi'_0=const,.
\end{equation}
 Since, for this stationary line,
$\gamma_{stat}={x_{stat}}/{A_{stat}}={1}/{A}=-{6\xi}/{N}$,
$Y_{stat}=-6$, we have
\begin{equation}
\begin{array}{l}
{|\phi(t)|}^{\frac{2-N}{2}}=\pm \frac{(2-N)\sqrt{6|\xi|}}{6}\ln
|\frac{t-t_0}{t'-t'_0}|.
\end{array}
\end{equation}
This solution exists for $ 0<N<2$.\\

 3). Finally, if $\phi=\phi_0=const$, $A=\frac{{N
\phi_0}^N}{1-6\xi{\phi_0}^N}$, again by using the combination
$\gamma_{stat}=\frac{x_{stat}}{A_{stat}}=\frac{1-6\xi{\phi_0}^N}{{N
\phi_0}^N}$ and substituting it in the expression of $\phi(t)$ in
eq.(\ref{eq:point2in1}), we get the same behavior, namely,
\begin{equation}
{|\phi(t)|}^{\frac{2-N}{2}}=\pm
\frac{(2-N)\sqrt{|\frac{1-6\xi{\phi_0}^N}{{\phi_0}^N}}|}{6}\ln
|\frac{t-t_0}{t'-t'_0}|.
\end{equation}
 Thus $\phi(t)\rightarrow\pm\infty$ for
$t\rightarrow t_0$ or $t\rightarrow\infty$ and contradicts  the
initial assumption that $\phi=\phi_0=const$.

    We therefore conclude that  the stationary line reduces to a
stationary point, $A=-\frac{N}{6\xi}$.
\\
\\
We can now write the expression for the quantity $G_N$ using the
coordinates of the stationary point,  $A={B'(\phi)\phi}/{(1-6 \xi
B(\phi))}={N  \phi^N}/{(1-6 \xi  \phi^N)}={N  }/{(\phi^{-N}-6 \xi
)}$, we find that $\phi^N=A/(N+6 \xi A)$ leading to
$G_N={6}/{8\pi(1-6 \xi \phi^N)}=6(N+6 \xi A)/8\pi N$ which is
positive when $\xi>0, A<-N/6\xi$ or $\xi<0, A>-N/6\xi$.
 \\
 \\
  Since the stationary line reduces to a stationary point for which $A=-N/6\xi$ {($\phi(t)\rightarrow\infty$)},
  it follows that
$G_N\rightarrow 0$. We also note that the vacuum solution does not
contain parameters, $\omega$,n and N.\\


%
\textbf{2.}~~~~~~~ $x=0, y=0, z=1,
A=-\frac{b+1}{6\xi}=-\frac{N}{6\xi}, \Omega=0$\\

The eigenvalues are,
\begin{eqnarray}
{\lambda}_1=\frac{b-1}{b+1}=\frac{N-2}{N},~~
{\lambda}_2=\frac{5+5b-c}{b+1}=5-\frac{n}{N},~~
{\lambda}_3=2-3\omega,~~ {\lambda}_4=1.
\end{eqnarray}
This point is unstable because ${\lambda}_4$ is positive for any
$\xi$, $\omega$. We can compute $a(t)$  by noting that $Y_{stat}=0$,
\begin{eqnarray}
a(t)=a_0{|t-t_0|}^{\frac{1}{2}}
\end{eqnarray}
As for the expression for $\phi(t)$, we use the following
combination of autonomous variables at the stationary point
\begin{equation}
\begin{array}{l}
\frac{z_{stat}}{6\xi A_{stat}}=\frac{\dot{\phi}}{\phi H}=\beta_{stat},\\
\\\phi(t)={\phi}_0{|t-t_0|}^{\frac{\beta_{stat}}{2-\frac{Y_{stat}}{6}}}.
\end{array}
\end{equation}
where $  \beta\equiv \frac{z}{6\xi A}=\frac{\dot{\phi}}{\phi H}$ and
we have made use of the fact that  $\beta_{stat}=const$. For the
vacuum stationary point under consideration,
$\beta_{stat}=\frac{z_{stat}}{6\xi
A_{stat}}=-\frac{1}{b+1}=-\frac{1}{N}$ which finally gives
\begin{equation}
\begin{array}{l}
\phi(t)={\phi}_0{|t-t_0|}^{-\frac{1}{2N}}.
\end{array}
\end{equation}

 Consistency analysis shows that this solution exists for $N>2$ and
$n<5N$, so it never coexists with the solution discussed above under
point {\bf 1}. They both correspond to the situation  in which the
scalar field potential is negligible.
Power indexes in this solutions do not depend on $\xi$, $\omega$,
$n$, however the function $\phi(t)$ contains dependence on $N$. We
also note that since $A=-N/ 6\xi$  {($\phi(t)\rightarrow\infty$ for
$t\rightarrow t_0)$} in the present case,
 $G_N$ vanishes asymptotically at the stationary point.\\

\textbf{3.}~~~$x=0, y=0, z=-1+3\omega,~~
A=-\frac{b+1}{6\xi}=-\frac{N}{6\xi},~~ \Omega=2-3\omega$\\

The corresponding eigenvalues are,
\begin{eqnarray}
{\lambda}_1&=&\frac{(1-b)(1-3\omega)}{b+1},\nonumber\\
{\lambda}_2&=&\frac{c(1-3\omega)+3(b+1)(1+\omega)}{b+1} <0,\nonumber\\
 &&\text{for $c>3(b+1)$}, {\omega}_0<\omega<1,\nonumber\\
{\lambda}_3&=&-2+3\omega <0, \text{    for $\omega <\frac{2}{3}$},\nonumber\\
{\lambda}_4&=&-1+3\omega <0, \text{    for $\omega <\frac{1}{3}$}.
\end{eqnarray}
where ${\omega}_0=-\frac{3(b+1)+c}{3(b+1-c)}>\frac{1}{3}$ for $b>0$, $c>3(b+1)$.

As regions where ${\lambda}_2$ and ${\lambda}_{4}$ are negative do
not intersect,  this point is unstable (either a saddle or a
repulsive node).

For obtaining $a(t)$, $\phi(t)$, we note that $Y_{stat}=0$,
$\beta=\frac{1-3\omega}{b+1} = \frac{1-3\omega}{N}$. This tells us that,
\begin{eqnarray}
a(t)&=&a_0{|t-t_0|}^{\frac{1}{2}},\\
\phi(t)&=&{\phi}_0{|t-t_0|}^{\frac{1-3\omega}{2N}}.
\end{eqnarray}
The time dependence of the matter density we obtain using its continuity equation
\begin{eqnarray}
\rho(t)=\rho_0{|t-t_0|}^{\frac{3(1+\omega)}{\frac{Y_{Stat}}{6}-2}}=\rho_0{|t-t_0|}^{-\frac{3(1+\omega)}{2}}
\end{eqnarray}
Power indexes of this solution do not contain parameters $\xi$, $n$ and
they depend on $\omega$, $N$.
\\
It follows from the definition of $\Omega$ that it must be positive
  to ensure  that $\rho>0$ and $G_N>0$. Since $\Omega=2-3\omega$ in the
  present case, we should have $\omega<2/3$ to avoid a pathological
  situation. We also note that since $A=-N/6\xi$ $(\phi(t)\to \infty)$,
  the effective Newtonian constant $G_N$ vanishes at the stationary point. \\

 \textbf{4.}
\begin{eqnarray}
x&=&0, y=\frac{(b+1)(c(1-3\omega)+3(b+1)(\omega+1))}{2c^2},\nonumber\\
z&=&\frac{3(b+1)(\omega+1)}{c}, A=-\frac{b+1}{6\xi}=-\frac{N}{6\xi},\\
\Omega&=&\frac{2c^2-(b+1)(3(\omega+1)(b+c+1)+4c)}{2c^2}\nonumber
\end{eqnarray}
The corresponding eigenvalues are,
\begin{eqnarray}
{\lambda}_1&=&\frac{3(\omega+1)(b-1)}{c},\nonumber\\
{\lambda}_{2,3}&=&\frac{3(\omega+1){(b+1)}^2+3c(\omega-1)(b+1))}{4c(b+1)}
\pm\frac{\sqrt{ (b+1)(f_1(b,c)\omega + f_2(b,c)+9{\omega}^2 f_3(b,c))}}{4c(b+1)},\nonumber\\
{\lambda}_4&=&\frac{3(\omega+1)(1+b)}{c},
\end{eqnarray}
where
\begin{eqnarray*}
 f_1(b,c)&=&-210c^2(b+1)+162(b^3+1)+192b^2 c
 +192c+486 b(b+1)+384bc+48c^3,\\
 f_2(b,c)&=&81(1+b^3)+(243b+17c^2)(1+b)
 +174c(1+b^2)-16c^3+348bc,\\
 f_3(b,c)&=&(1+b)(b+1+c)(9(b+1)-7c).
\end{eqnarray*}
and ${\lambda}_1$, ${\lambda}_4$ are positive for $-1<\omega\leqslant 1$ therefore this stationary point is unstable for any $\xi$. \\

Analogous to previous points we get,
\begin{eqnarray*}
Y&=&\frac{3(3(b+1)(\omega+1)+c(1-3\omega))}{c}
=\frac{3(3 N(\omega+1)+n(1-3\omega))}{n},\\
\beta &=&-\frac{3(\omega+1)}{c}=-\frac{3(\omega+1)}{n}.
\end{eqnarray*}
then,
\begin{eqnarray}
\label{eq:rho11}
a(t)&=&a_0{|t-t_0|}^{-\frac{2n}{3(N-n)(\omega+1)}},\nonumber\\
\phi(t)&=&{\phi}_0{|t-t_0|}^{\frac{2}{(N-n)}},\nonumber\\
\rho(t)&=&{\rho}_0{|t-t_0|}^{\frac{2n}{N-n}}.
\end{eqnarray}
We note that power indices of this solution depend on $\omega$, $N$,
$n$ and do not depend on  the coupling constant $\xi$.

    For $b+1=c$ ($N=n$), the power index
    of functions $a(t)$ and $\phi(t)$ is infinite
     and power-law solutions cease to exist. In this case the
     coordinates of the fixed point 4
     are, $x=0, y=2, z=3(\omega+1), A=-\frac{b+1}{6\xi},
     \Omega=-(3\omega+4)$.
    We note that $\Omega<0$ for $\omega\in[-1,1]$ and that from
    the definition of $\Omega$ in (\ref{eq:omega}), it follows
    that either $\rho>0$, $G_N<0$ or $\rho<0$, $G_N>0$
    which doesn't correspond to the real Universe. We
    also  find in this case that $Y_{stat}=12$,
    $\beta=-\frac{3(1+\omega)}{N}$. This tells
    us that
    ${\dot{H}}/{H^2}=0$ and ${\dot{\phi}}/{\phi}=\beta H_0$ which is
    equivalent to
\begin{eqnarray}
\label{eq:rho}
a(t)&=&a_0 e^{H_0(t-t_0)},\nonumber\\
\phi(t)&=&\phi_0 e^{\beta H_0(t-t_0)}=\phi_0 e^{-\frac{3 H_0(1+\omega)(t-t_0)}{N}},\nonumber\\
\rho(t)&=&{\rho}_0 e^{-3 H_0(1+\omega)(t-t_0)}.
\end{eqnarray}
As for the constants $H_0$ and $\rho_0$,
 we substitute (\ref{eq:rho}) in definitions (\ref{eq:omega}) of
    $y$, $\Omega$ taking into account that $N=n$, $y=2$ and
    $\Omega=-(3\omega+4)$. We then  have
\begin{eqnarray}
\label{eq:omega1} y&=&\frac{2V_0\phi^n}{H^2(1-6\xi\phi^N)}
=\frac{2V_0{\phi_0}^N e^{-3 H_0(1+\omega)(t-t_0)}}{{H_0}^2\left( 1-6\xi{\phi_0}^N e^{-3 H_0(1+\omega)(t-t_0)}\right) }\rightarrow -\frac{V_0}{3{H_0}^2 \xi}\nonumber\\
\Omega&=&\frac{2\rho}{H^2(1-6\xi\phi^N)}
=\frac{2\rho_0 e^{-3 H_0(1+\omega)(t-t_0)}}{{H_0}^2\left( 1-6\xi{\phi_0}^N e^{-3 H_0(1+\omega)(t-t_0)}\right) }\rightarrow -\frac{\rho_0}{3{H_0}^2{\phi_0}^N \xi}\nonumber\\
\end{eqnarray}
\\
\text{for $t\rightarrow t_0$ when $H_0>0$ (or for $t\rightarrow
\infty $ when $H_0<0$)}.
\\
 Therefore,
\be
 {H_0}^2=-\frac{V_0}{3\xi y}=-\frac{V_0}{6\xi}, \nonumber
\ee
where $\xi<0$, $V_0>0$ or $\xi>0$, $V_0<0$,
and
\be
\rho_0=-3\Omega{H_0}^2{\phi_0}^N \xi=-\frac{V_0{\phi_0}^N(3\omega+4)}{2}.\nonumber
\ee
\\
  We next consider the behavior of quantities $T=T_{\phi}+T_{m}$ and $R=6T_{eff}=8\pi G_N T$
   for the case $N=n$.
   Substituting the so obtained exponential
   solution, $a(t)=a_0 e^{H_0(t-t_0)}$,
   $\phi(t)=\phi_0 e^{-\frac{3H_0(1+\omega)(t-t_0)}{N}}$ in (\ref{eq:Friedphi2}), we find
\begin{eqnarray}
\label{eq:teff}
T_{eff}&=&\frac{R}{6}=2{H_0}^2=const,\nonumber\\
G_N&=&\frac{6}{8\pi(1-6\xi{\phi_0}^N e^{-3H_0(1+\omega)(t-t_0)})}\propto e^{3H_0(1+\omega)(t-t_0)}\nonumber\\
T&=&T_\phi+T_m\nonumber\\
&=&e^{-3H_0(1+\omega)(t-t_0)}\left( {{\phi}_0}^N (4
V_0-27\xi(1+\omega){H_0}^2
+27\xi{(1+\omega)}^2{H_0}^2)+\rho_0(1-3\omega)\right) \propto
e^{-3H_0(1+\omega)(t-t_0)}
\end{eqnarray}
\text{for $t\rightarrow t_0$, $H_0>0$ (or $t\rightarrow\infty$, $H_0<0$)}.\\
\\
It is therefore clear from the aforesaid that $G_N(t)$ grows as an
exponent whereas $T(t)$ decreases with the same rate thereby leading
a constant  product $G_N T$. {A remark about the exponentially
expanding solution is in order. The solution though has features
similar to de Sitter solution but does not really qualify for a true
de Sitter as $G_N$ is not constant in this case. We shall say more
about this point in the discussion of the vacuum solution to
follow.}
\\

\textbf{5. Vacuum solution}\\

\begin{figure*} \centering
\begin{center}
$\begin{array}{c@{\hspace{0.4in}}c}
\multicolumn{1}{l}{\mbox{}} &
        \multicolumn{1}{l}{\mbox{}} \\ [0.0cm]
\epsfxsize=3.2in \epsffile{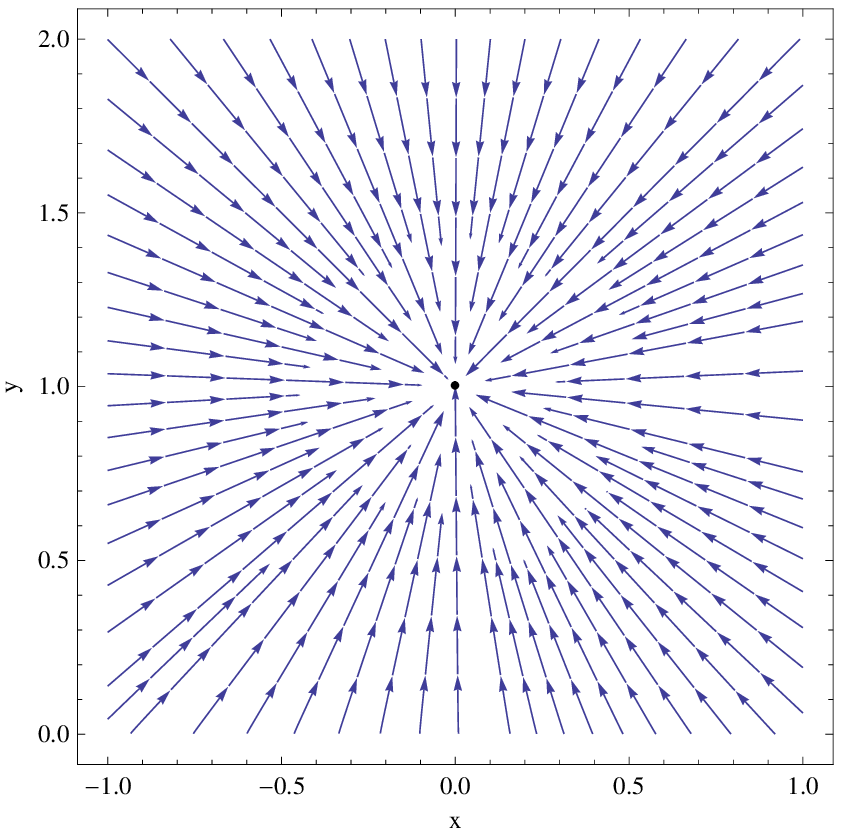} &
        \epsfxsize=3.2in
        \epsffile{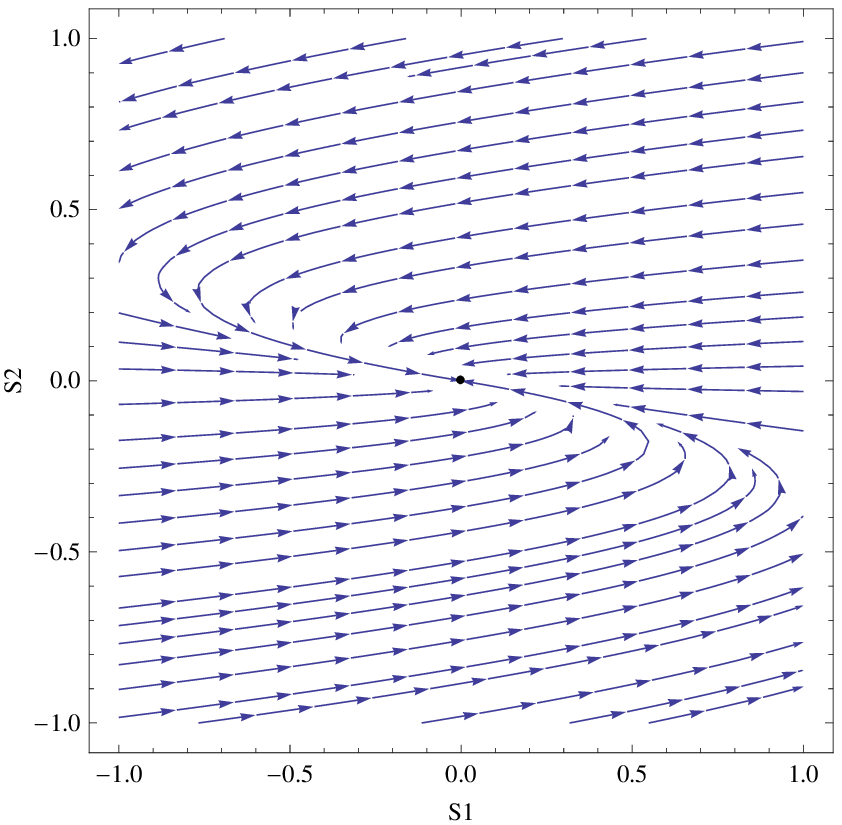} \\ [0.20cm]
\mbox{\bf (a)} & \mbox{\bf (b)}
\end{array}$
\end{center}
\caption{ The left panel (a) shows the Phase portrait for negative
coupling constant $\xi=-1/2$ for the fixed point 5, the fixed point
is an attractive node.The right panel (b) displays  the Phase
portrait for positive coupling constant $\xi=0.3961$ ($N=2,n=7$) for
de sitter which is an attractive focus.} \label{vds}
\end{figure*}

 This solution corresponds to the
following fixed point,
\begin{eqnarray}
x=0,~ y=\frac{5+5b-c}{b+1+c},~ z=-\frac{2(2+2b-c)}{b+1+c},~~
A=-\frac{b+1}{6\xi}=-\frac{N}{6\xi},~~ \Omega=0.
\end{eqnarray}
The corresponding eigenvalues in this case are given by,
\begin{eqnarray}
{\lambda}_1&=&-\frac{2((b-1)(2(b+1)-c))}{(b+1)(b+1+c)}<0,
\text{  for $b>1$ and $c<2(b+1)$}, \nonumber\\
{\lambda}_2&=&-\frac{2(2+2b-c)}{b+1+c}<0,
 \text{  for $c<2(b+1)$},\nonumber\\
{\lambda}_3&=&-\frac{5+5b-c}{b+1}<0, \text{  for
$c<5(b+1)$}.\nonumber\\
{\lambda}_4&=&-\frac{(b+1)(3(b+1)(\omega+1)+c(7+3\omega))-2c^2}{(b+1)(b+1+c)}<0,\nonumber\\
&&\text{  for $c<2(b+1)$ when $\omega\in[-1;1]$},
\end{eqnarray}
 The negativity  of eigenvalues for this vacuum point show that it
is stable for $c<2(b+1)$. Indeed, it is an attractive node
 for $b>1$ and $c<2(b+1)$ for any numerical values of $\xi$ and $\omega$, see Fig.\ref{vds}(a).\\
We note that,
\begin{eqnarray*}
Y&=&\frac{6c(5+5b-c)}{(b+1+c)(b+1)}=\frac{6 n(5 N-n)}{(N+n)N}, \\
\beta&=&\frac{2(2+2b-c)}{(b+1+c)(b+1)}=\frac{2(2 N-n)}{(N+n)N}
\end{eqnarray*}
which gives rise to the following expressions for $a(t)$ and
$\phi(t)$,
\begin{eqnarray}
\label{eq:phi1}
a(t)&=&a_0{|t-t_0|}^{\frac{(N+n)N}{(2 N-n)(N-n)}},\nonumber\\
\phi(t)&=&{\phi}_0{|t-t_0|}^{\frac{2}{N-n}}
\end{eqnarray}
This solution contains parameters $N$, $n$ but is  independent of
$\xi$ and $\omega$. We note that power indexes in (\ref{eq:phi1})
are negative for $N<n<2N$ and, therefore, $a(t)$, $\phi(t)$
  diverge leading to "Big Rip" singularity at $t=t_0$. This result is generalization
   for
  the analogous vacuum
  solution $a(t)=a_0{|t-t_0|}^{\frac{2 (\xi(2+n)-1)}{ \xi(n-2)(n-4)}}$, $\phi(t)={\phi}_0{|t-t_0|}^{\frac{2}{2-n}}$
  obtained in the case of $N=2$ \cite{Polarski,Dunsby1}.

    Let us further investigate the nature of the fixed point.   For $b+1=c$ ($N=n$) and also for $2(b+1)=c$ ($2N=n$)
    power indexes
      of functions $a(t)$ and $\phi(t)$ diverge  and
      power-law solutions should transform into exponential ones.
      Indeed, the coordinates of the fixed point in these
      cases are given by, $x=0, y=2, z=-1, A=-\frac{N}{6\xi}, \Omega=0$
       and $x=0, y=1, z=0, A=-\frac{N}{6\xi}, \Omega=0$
        respectively for $N=n$ and $2N=n$.  In the first case, $Y=12$,
         $\beta=\frac{1}{N}$  whereas  $Y=12$,
         $\beta=0$ in the second. In case of $N=n$, we find that, $\frac{\dot{H}}{H^2}=0$, $H=H_0=const$ and
\begin{eqnarray}
\label{eq:phi2}
a(t)&=&a_0 e^{H_0(t-t_0)},\nonumber\\
\phi(t)&=&\phi_0 e^{\beta H_0(t-t_0)}=\phi_0 e^{\frac{H_0(t-t_0)}{N}}.
\end{eqnarray}
We can find out $H_0$, using the definition of the coordinate $y$
from (\ref{eq:omega}),
\begin{eqnarray}
\label{eq:phi3}
y&=&\frac{2V_0\phi^n}{H^2(1-6\xi\phi^N)}=\frac{2V_0{\phi_0}^N e^{H_0(t-t_0)}}{{H_0}^2\left( 1-6\xi{\phi_0}^N e^{H_0(t-t_0)}\right) }\nonumber\\
&&\rightarrow -\frac{V_0}{3{H_0}^2 \xi}~~for~~t\rightarrow \infty~,
(H_0>0)~ or~ t\rightarrow t_0~ (H_0<0).
\end{eqnarray}

Thus, ${H_0}^2=-\frac{V_0}{3\xi y}=-\frac{V_0}{6\xi}$ where
$\xi<0$, $V_0>0$ or $\xi>0$, $V_0<0$.\\

In case of $n=2N$, the fixed point under consideration does not lead
to any physically
 admissible regime ( our
 numerical work shows that all trajectories in this case lead to oscillations near the
 minimally coupled field solution).
It is interesting to note that in the absence of  the standard
curvature term in the action ($F(\phi)R=\phi^N R$), a family of de
Sitter solutions  exists \cite{Kamen} for
 arbitrary values of field $\phi$.
 As for $n=N$, the solution corresponds to $\dot{H}=0$
and $w_{eff}=-1$ taken usually as the definition for de-Sitter
solution tacitly assuming that Newtonian gravitational constant is a
true constant of nature. In case, $\phi$ is constant, the constancy
of $G_N$  is trivially satisfied. However, in the model under
consideration, we have an interesting vacuum solution with
$\dot{H}=0$ and an exponentially expanding solution for $\phi$ which
corresponds to an exponentially decreasing/increasing (depending
upon the sign of $H_0$) the effective Newtonian gravitational
constant, $G_N$. The true de-Sitter corresponds to $\dot{H}=0$ and
$\phi=const$ implying $G_N=const$. Let us note that in this case,
choosing $\phi_0=0$, one might think  to obtain de Sitter solution
but the latter implies $H_0=0$. Clearly, this solution does not
qualify for a genuine de Sitter.

In fact, the true de-Sitter solution is not captured by the
autonomous variables, we have used. In this case, the combination
$4x+z^2$ appearing in the denominator of autonomous system vanishes
thereby telling us that autonomous variables used here are not
suitable for the investigation of de-Sitter solution. We shall
discuss this special case in a separate section to follow.
\begin{figure}
  \begin{center}
   $\begin{array}{c@{\hspace{0.4in}}c}
\epsfxsize=3.2in \epsffile{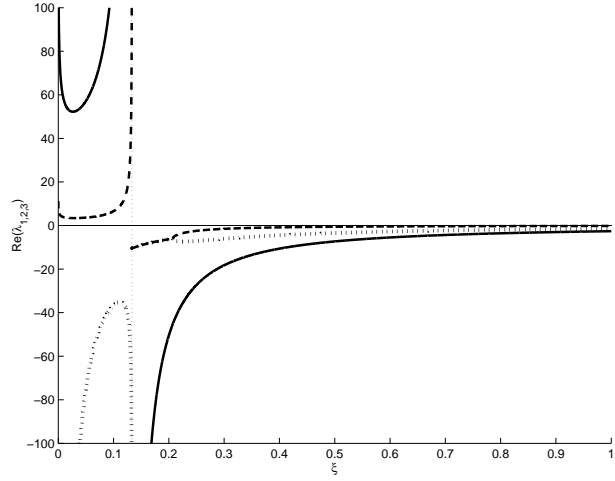}
\end{array}$
 \end{center}
 \caption{\small  This figure shows the real parts of eigen values for $N=2$
 and $n=5$ versus $\xi$. The figure shows that the real part of two of the three eigen values
 is positive for $\xi<0.1333$ making the de Sitter solution unstable.
 Beyond the quoted value of $\xi$, real parts of all
 the eigen values remain negative. The two eigen
 values are complex conjugate of each other (their real parts coincide) in case of
 $,0.1333<\xi\leq0.2068$ and we see only two curves. Beyond
 $\xi=0.2068$, the eigen value curve splits into two. }
 \label{evalues}
\end{figure}

\section{Investigation of $de$-Sitter solution: its existence and stability}
\label{stability}

  As mentioned earlier, our autonomous system is not suitable
for the study of de-Sitter solution corresponding to, $x=0$, $y=1$,
$z=0$, $A=-\frac{n}{12\xi}$, $\Omega=0$ as the combination, $4x+z^2$
appearing in the denominator of dynamical system  vanishes
identically. Though the formalism of dynamical systems using
autonomous variables is very powerful for the study of fixed points
and their stability but it might miss some particular dynamical
feature of the original system. In such a case we should go back to
the original description to capture the same and this is what we
shall do in the discussion to follow.

In this special, we go back to the original variables and make use
of the system of equations (\ref{eq:Friedphi}),
(\ref{eq:Friedphi2}), (\ref{eq:KGphi}). Substituting,
$\dot{H}=\dot{\phi}=\ddot{\phi}=\rho=0$ and $R=12{H_0}^2$, for the
de-Sitter solution in these equations we find,

\begin{eqnarray}
\label{eq:b14} {H_0}^2(1-6\xi {\phi_0}^N)=2 V_0 {\phi_0}^n,~~~
 6 {H_0}^2\xi N {\phi_0}^{N-1}+V_0 n {\phi_0}^{n-1}=0 ,
\end{eqnarray}
which gives important information for solution under consideration,
\begin{eqnarray}
\label{eq:b15} H_0^2=-\frac{V_0 n{\phi_0}^{n-N}}{6\xi  N},~~~
{\phi_0}^N=\frac{n}{6\xi (n-2N)}.
\end{eqnarray}
Clearly, this solution does not exist for $n>2N$ in general for an
even $n$. We also note that for de-Sitter solution , the effective
Newtonian gravitational constant, $G_N={6}/{8\pi(1-6\xi
B(\phi))}=3(2N-n)/8\pi N$ is positive only if $n<2N$.

      The system of equations  (\ref{eq:Friedphi}), (\ref{eq:Friedphi2}), (\ref{eq:KGphi}) for $\rho=0$
      which is of interest to us reduces to,
\begin{eqnarray}
\label{eq:ds1} \dot H (1-6\xi
B+9{\xi}^2{B'}^2)&=&3{\dot{\phi}}^2(\xi B''-1)-3\xi B'(4H \dot{\phi}+V'+6\xi B' H^2),\nonumber\\
\ddot{\phi}(1-6\xi B+9\xi^2{B'}^2)&=&-(3H \dot{\phi}+V')(1-6\xi
B)-\xi B'(-3\dot{\phi}^2+12V+9\xi(3H \dot{\phi} B'+{\dot{\phi}}^2
B'')).
\end{eqnarray}
\begin{figure*} \centering
\begin{center}
$\begin{array}{c@{\hspace{0.4in}}c} \multicolumn{1}{l}{\mbox{}} &
        \multicolumn{1}{l}{\mbox{}} \\ [0.0cm]
\epsfxsize=3.2in \epsffile{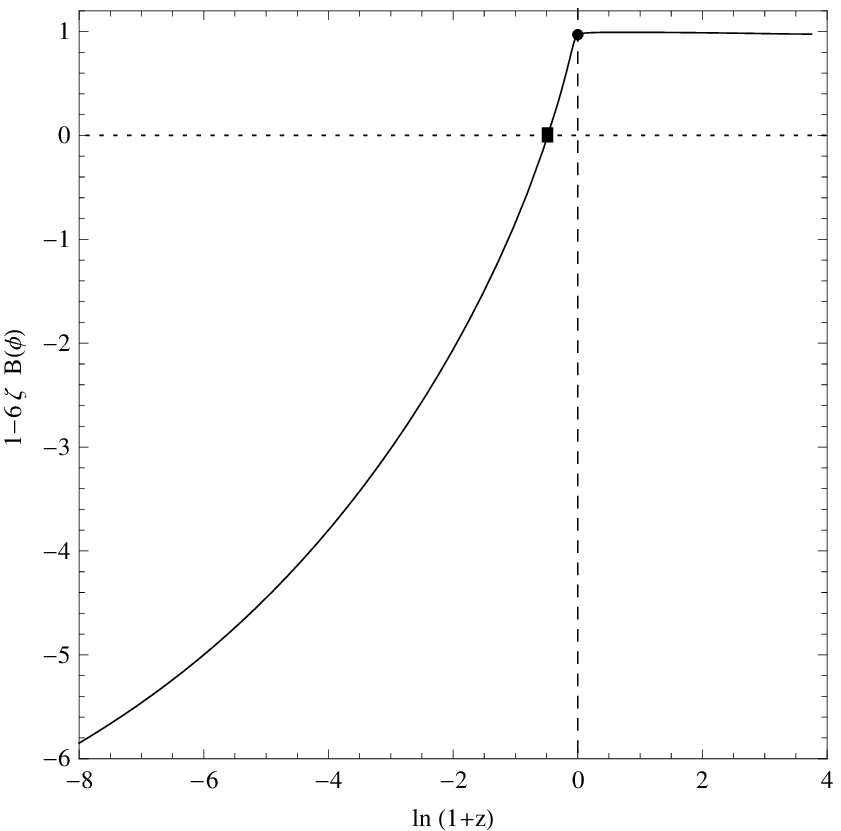} &
        \epsfxsize=3.2in
        \epsffile{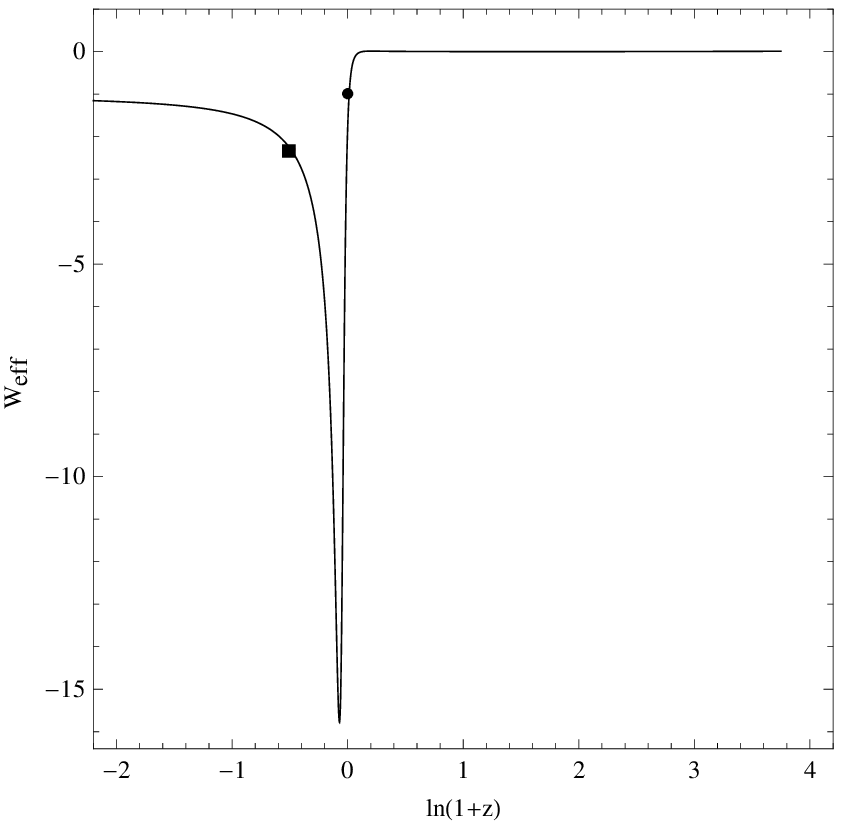} \\ [0.20cm]
\mbox{\bf (a)} & \mbox{\bf (b)}
\end{array}$
\end{center}
\caption{\small The left panel (a) shows the evolution of $(1-6\xi
B(\phi))$ versus redshift z, where $G_N=6/{8\pi(1-6\xi B(\phi))}$.
The right panel (b) shows the evolution of $w_{eff}$ versus redshift
z. Both the figures correspond to the case of non-minimal coupling
with $N=4$, $n=9$, $\xi= 1/2$. We had chosen appropriate initial
conditions, $H=2.034, \phi=-0.300,\dot{\phi}=-0.001$ and
$\xi=0.5000$ (corresponding to $w_m=0$ initially) to obtain
$w_{eff}\simeq -1$ at the present epoch. The black dot in both the
figures designates the present epoch which occurs in the regime of
$G_N>0$, the effective Newtonian constant changes sign thereafter in
future. The square marks the epoch where $G_N$ turns negative.}
\label{gw}
\end{figure*}

In order to investigate the stability of the de-Sitter solution, we
 consider small perturbations $\mu$ $\&$ $\nu$ around this
background: $H=H_0+\mu$ and $\phi={\phi}_0+\nu$ in the dynamical
system (\ref{eq:ds1}). Consequently, we find the evolution equations
for perturbations
\begin{eqnarray}
\label{eq:ds2} &&\mu(36{\xi}^2  N^2 H_0 {{\phi}_0}^{2N-2})
+\dot{\mu}(1-6\xi  {{\phi}_0}^N+9{\xi}^2 N^2{{\phi}_0}^{2N-2})+\nu(3\xi  V_0 Nn(N+n-2){{\phi}_0}^{N+n-3}\nonumber\\
&&+36{\xi}^2  N^2(N-1){H_0}^2{{\phi}_0}^{2N-3})+\dot{\nu}(12\xi  N
H_0 {{\phi}_0}^{N-1})=0,
\end{eqnarray}
\begin{eqnarray}
\label{eq:ds3} &&\nu(V_0
n(n-1){{\phi}_0}^{n-2}(1-6\xi{\phi_0}^N)-6\xi  V_0
Nn{{\phi}_0}^{N+n-2}+12\xi  V_0
N(N+n-1){{\phi}_0}^{N+n-2})\nonumber\\
&&+\dot{\nu}(3H_0 (1-6\xi {{\phi}_0}^N)+27{\xi}^2 N^2
H_0{{\phi}_0}^{2N-2}) +\ddot{\nu}(1-6\xi {{\phi}_0}^N+9{\xi}^2
N^2{{\phi}_0}^{2N-2})=0.
\end{eqnarray}
In order to cast equations (\ref{eq:ds2}), (\ref{eq:ds3}) in a
simple form, we introduce the following notations
\begin{eqnarray}
D&=&36{\xi}^2  N^2 H_0 {{\phi}_0}^{2N-2},\nonumber\\
E&=&1-6\xi {{\phi}_0}^N+9{\xi}^2 N^2{{\phi}_0}^{2N-2},\nonumber\\
F&=&3\xi  V_0 Nn(N+n-2){{\phi}_0}^{N+n-3}+36{\xi}^2  N^2(N-1){H_0}^2{{\phi}_0}^{2N-3},\nonumber\\
G&=&12\xi  N H_0 {{\phi}_0}^{N-1},\nonumber\\
D_1&=&V_0 n(n-1){{\phi}_0}^{n-2}(1-6\xi {{\phi}_0}^N)-6\xi  V_0 Nn{{\phi}_0}^{N+n-2}+12\xi  V_0 N(N+n-1){{\phi}_0}^{N+n-2},\nonumber\\
E_1&=&3H_0 (1-6\xi {{\phi}_0}^N)+27{\xi}^2 N^2 H_0{{\phi}_0}^{2N-2},\nonumber\\
F_1&=&1-6\xi {{\phi}_0}^N+9{\xi}^2 N^2{{\phi}_0}^{2N-2}\nonumber\\
\end{eqnarray}
and make use of new variables $s_1=\mu$, $s_2=\nu$, $s_3=\dot{\nu}$.
The system of equations (\ref{eq:ds2}) and (\ref{eq:ds3}) then
acquires a simple form,
\begin{eqnarray}
&&D s_1+E \dot{\mu} +F s_2+G s_3=0,\\
&&D_1 s_2+E_1 s_3+F_1 \ddot{\nu}=0.
\end{eqnarray}
Taking derivative of $s_1$, $s_2$, $s_3$ with respect to time we get the system of equations,
\begin{equation}
\label{eq:matrix}
\left(\begin{array}{c}
\dot{s_1}\\
\dot{s_2}\\
\dot{s_3}
\end{array}\right)
=
\left(\begin{array}{ccc}
-\frac{D}{E}&-\frac{F}{E}&-\frac{G}{E}\\
 0&0&1\\
0&-\frac{D_1}{F_1}&-\frac{E_1}{F_1}
\end{array}\right)
\left(\begin{array}{c}
s_1\\
s_2\\
s_3
\end{array}\right)
\end{equation}
which we have studied numerically. The numerical investigation of
this system for $b=N-1=1,2,3,4,5,6$ and $c=n=1,2,3,4,5,6,7,8,9$
shows that the de Sitter solution is stable when (we assume $V_0=1$)
 $2N+1\leqslant n =5,7,9,...,$($N=2,4,6,...$) and for positive
values of the coupling, $\xi\ > {\xi}_0>0$, where ${\xi}_0$
depends on $b$, $c$.
 We hereby quote few cases which show how the nature of the fixed points
 crucially depends upon numerical values of the coupling $\xi$:\\

$\bullet$ For b=1 (N=2) and c=n=5, the fixed point is\\
saddle if~ $0<\xi\leq \xi_0, \xi_0\simeq0.1333$,\\
attractive focus if~$\xi_0<\xi\leq 0.2068$,\\
attractive node if ~$0.2069\leq \xi\leq 1$;\\

$\bullet$ For b=1 (N=2) and c=n=7, the fixed point is\\
saddle if ~$0<\xi\leq \xi_0, \xi_0\simeq0.0952$,\\
attractive focus if~$\xi_0<\xi\leq0.3999$,\\
attractive node if ~$0.4000\leq \xi\leq 1$;\\

$\bullet$ For b=1 (N=2) and c=n=9, the fixed point is\\
saddle if ~$ 0<\xi \leq \xi_0, \xi_0\simeq 0.0740$,\\
attractive focus if~$\xi_0<\xi\leq 1$;\\

$\bullet$ For b=3(N=4) and c=n=9, the fixed point is\\
saddle if ~$0<\xi<\xi_0, \xi_0\simeq0.0009$,\\
attractive focus if~$\xi_0<\xi\leq0.0014$,\\
attractive node if ~$0.0015\le \xi\leq 1$.\\
\begin{figure}
   \begin{center}
     \includegraphics{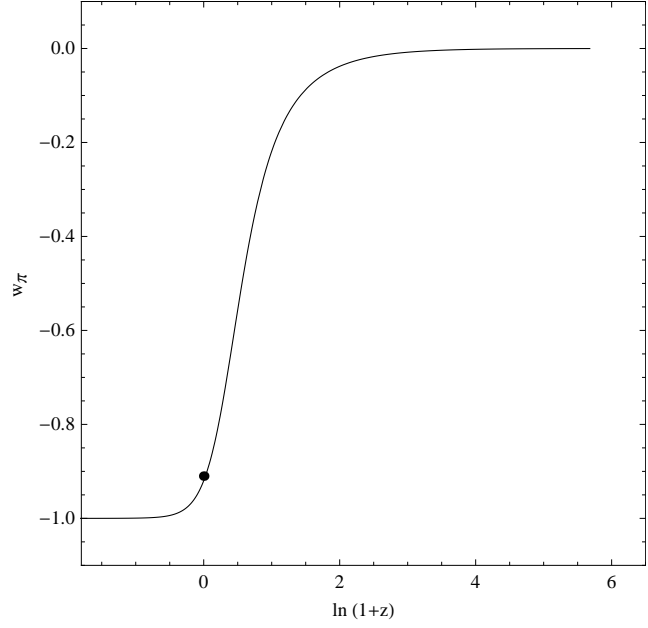}
 \end{center}
\caption{\small This figure shows the evolution of the equation of
state parameter $w_{\pi}$ versus redshift z for
 a scalar field $\pi$ in galileon model with Friedmann equation (\ref{eq:Fried1}). The initial
 conditions were chosen carefully allowing us to be in the region of phase space where de Sitter is a
  stable fixed point\cite{Gannouji:2010au}. The black dot on the curve designates the present epoch.}
 \label{wpi}
\end{figure}
The nature of eigen values of the perturbation matrix corresponding
to eq.(\ref{eq:matrix}) crucially depends upon the numerical values
of $b$, $c$ and $\xi$. For instance, the eigen values for the fourth
point, $(N,n,\xi)=4,9,0.0012$ are, $-47990.1409, -2286.0463 \pm
1703.0817i$ which shows that de Sitter solution is an attractive
focus. We numerically investigated the region of stability of the
solution, see Fig.\ref{evalues}. We did not find stable de Sitter
solution in the range of $n<2N$ which has important
implication for late time cosmology. \\

The stable de Sitter solution exists in the region where the
effective Newtonian constant $G_N$ is negative which means that
graviton is ghost thereby leading to instability. Before approaching
the attractor, the system passes through a phantom phase and
parameters in the theory can easily be adjusted such that we obtain
the observed value of equation of state parameter, $w_{eff} \simeq
-1$ at present with $G_N>0$ followed by a brief phantom phase before
approaching the stable de Sitter fixed point ultimately pushing the
ghost dominated regime to future (see Fig.\ref{gw}). It is also
possible to set the phantom phase at the present epoch. We have
carefully managed to shift the ghost regime to future by adjusting
the parameters in the model. We should, however, admit that such a
model of transient dark energy suffers from ugly fine tuning
problem.


\begin{figure*}
\begin{center}
$\begin{array}{c@{\hspace{0.4in}}c}
\multicolumn{1}{l}{\mbox{}} &
        \multicolumn{1}{l}{\mbox{}} \\ [0.0cm]
\epsfxsize=3.2in
\epsffile{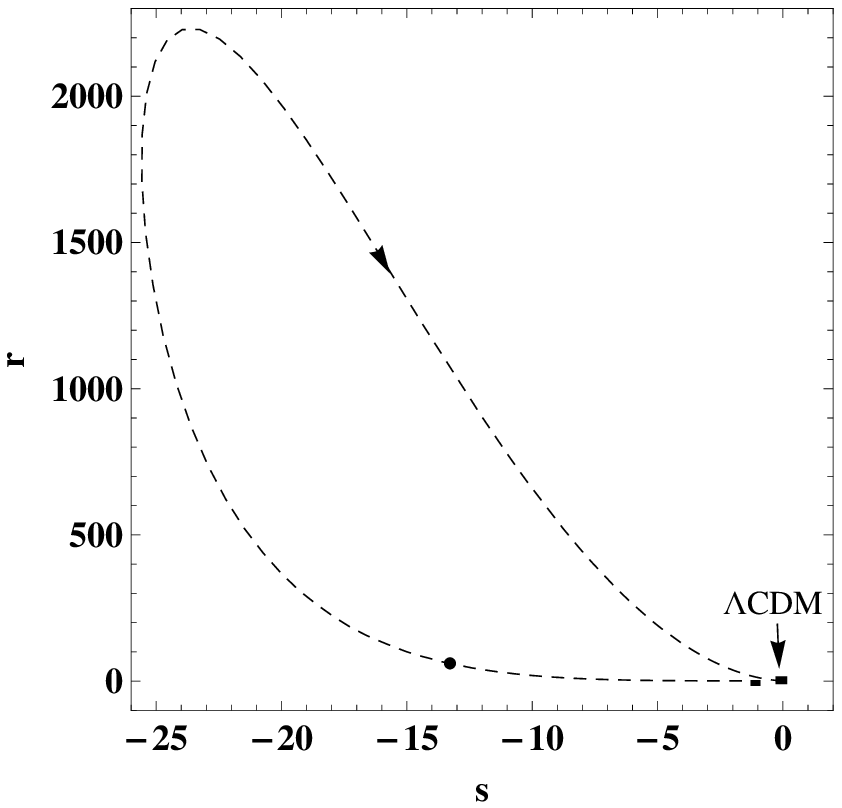} &
        \epsfxsize=3.2in
        \epsffile{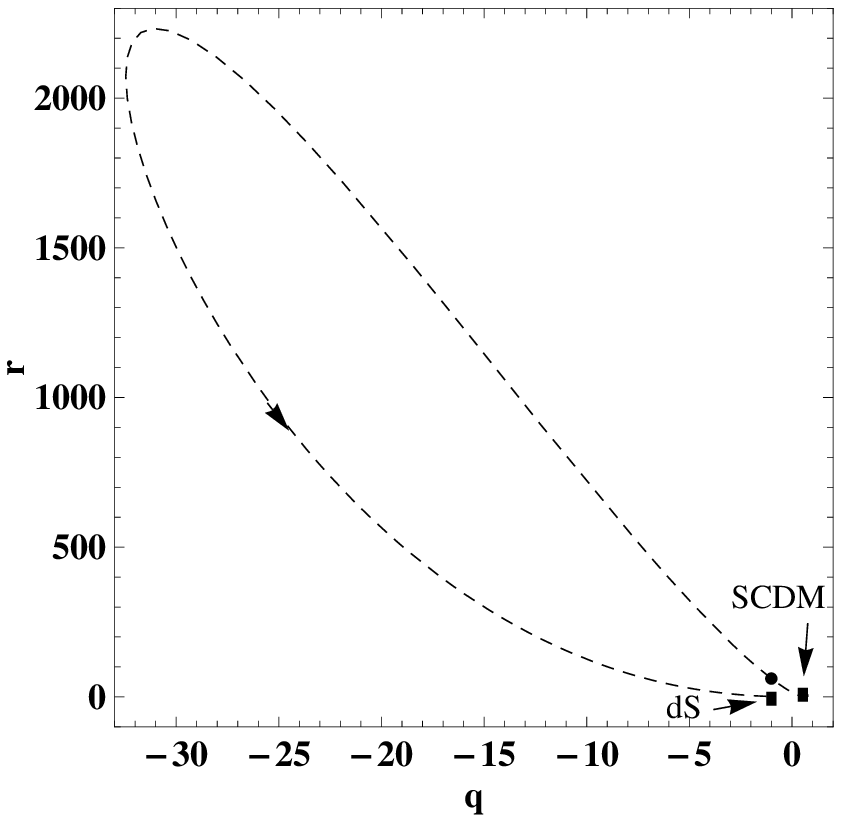} \\ [0.20cm]
\mbox{\bf (a)} & \mbox{\bf (b)}
\end{array}$
\end{center}
\caption{ In this figure, we plot the statefinder parameters for
non-minimal model. The left panel (a) shows the time evolution of
the statefinder pair $\lbrace r,s \rbrace$ . The model converge to
the fixed point ($r=1, s=0$).The right panel (b) shows the time
evolution of the statefinder pair $\lbrace r,q\rbrace$, the model
diverge at the point ($r=1,q=0.5$) which corresponds to matter
dominated universe and converge to the point ($r=1,q=-1$) which
corresponds to de Sitter expansion.The dark dots on the curves show
values $\lbrace r_0,s_0 \rbrace$ (left) and $\lbrace r_0,q_0
\rbrace$ at the moment when $w_{eff}\simeq-1$ from FIG. 3 (b)
(right). Black squares show ($r=1,s=-1,0$) (left) and ($r=1,q=-1,
0.5$) (right).}
  \label{stat0}
\end{figure*}
\begin{figure*} \centering
\begin{center}
$\begin{array}{c@{\hspace{0.4in}}c}
\multicolumn{1}{l}{\mbox{}} &
        \multicolumn{1}{l}{\mbox{}} \\ [0.0cm]
\epsfxsize=3.2in
\epsffile{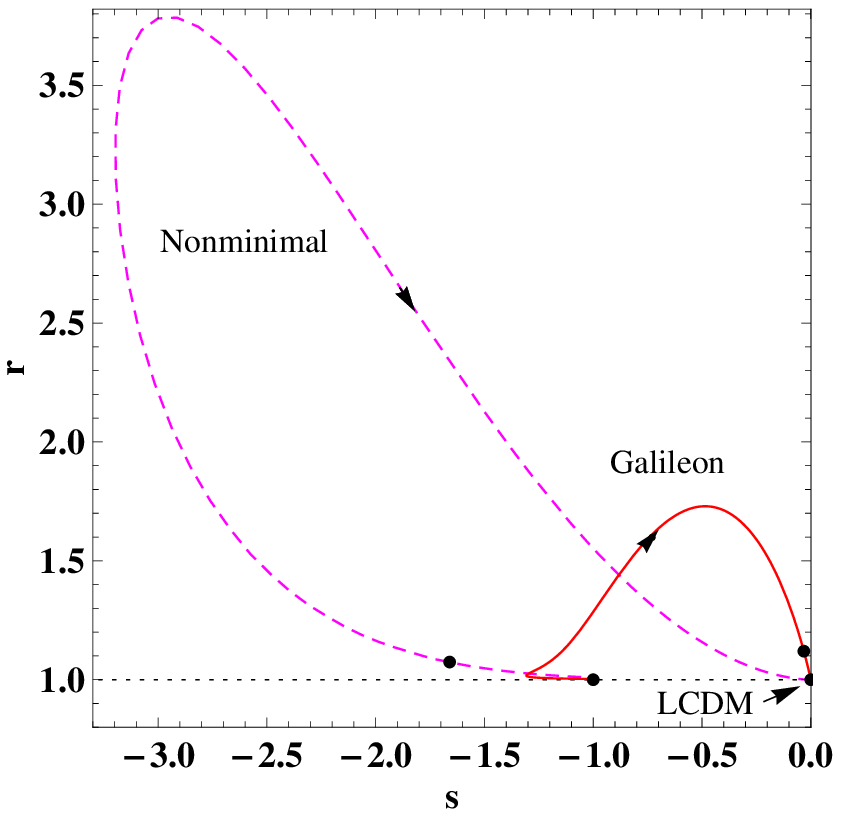} &
        \epsfxsize=3.2in
        \epsffile{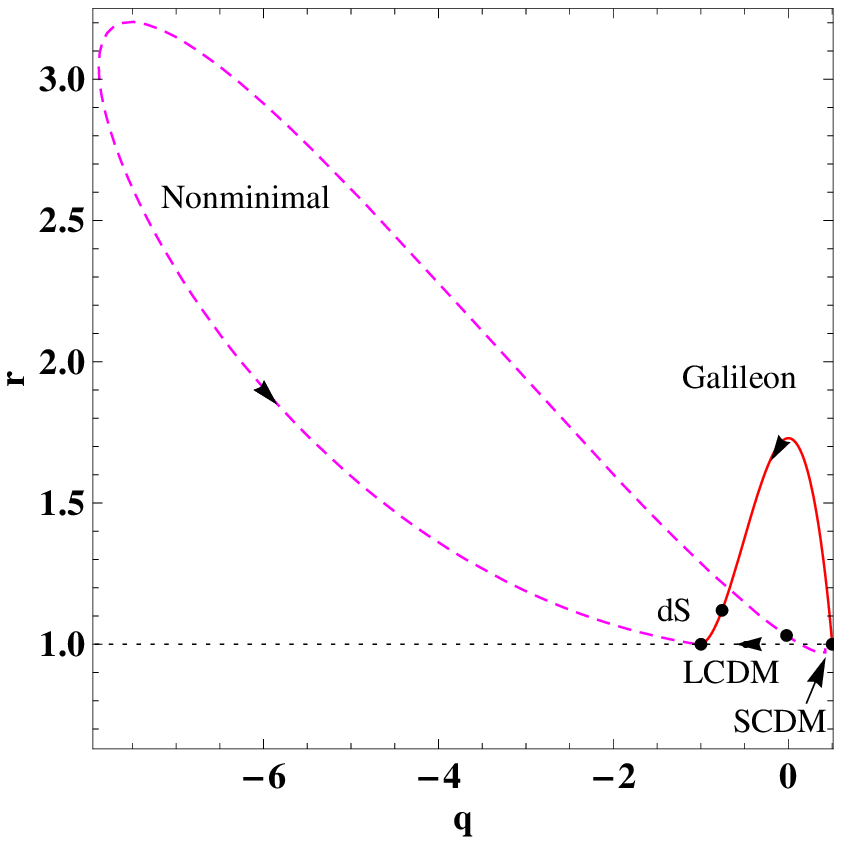} \\ [0.20cm]
\mbox{\bf (a)} & \mbox{\bf (b)}
\end{array}$
\end{center}
\caption{(color online) The left panel (a) shows the time evolution
of the statefinder pair $\lbrace r,s \rbrace$ for galileon (red) and
Non-minimal (magenta) models. Both galileon and non-minimal models
lie to the left of the $\Lambda$CDM fixed point ($r=1,s=0$). For
both the  models, $s$ increases to zero from -1, whereas $r$ first
increases from unity to a maximum value, then decreases to unity.
Both models converge to the fixed point ($r=1, s=0$) which
corresponds to $\Lambda$CDM.The right panel (b) shows the time
evolution of the statefinder pair $\lbrace r,q\rbrace$ for galileon
(red) and non-minimal (magenta) models. Both models diverge at the
same point ($r=1,q=0.5$) which corresponds to a matter dominated
universe (SCDM) and converge to the same point ($r=1,q=-1$) which
corresponds to the de Sitter expansion(dS).The dark dots on the
curves show current values $\lbrace r_0,s_0 \rbrace$ (left) and
$\lbrace r_0,q_0 \rbrace$ (right) for different dark energy models.
In all models, we have taken, $\omm = 0.3$  at the current epoch.}
 \label{stat1}
\end{figure*}
\begin{figure*}
\begin{center}
$\begin{array}{c@{\hspace{0.4in}}c}
\multicolumn{1}{l}{\mbox{}} &
        \multicolumn{1}{l}{\mbox{}} \\ [0.0cm]
\epsfxsize=3.2in
\epsffile{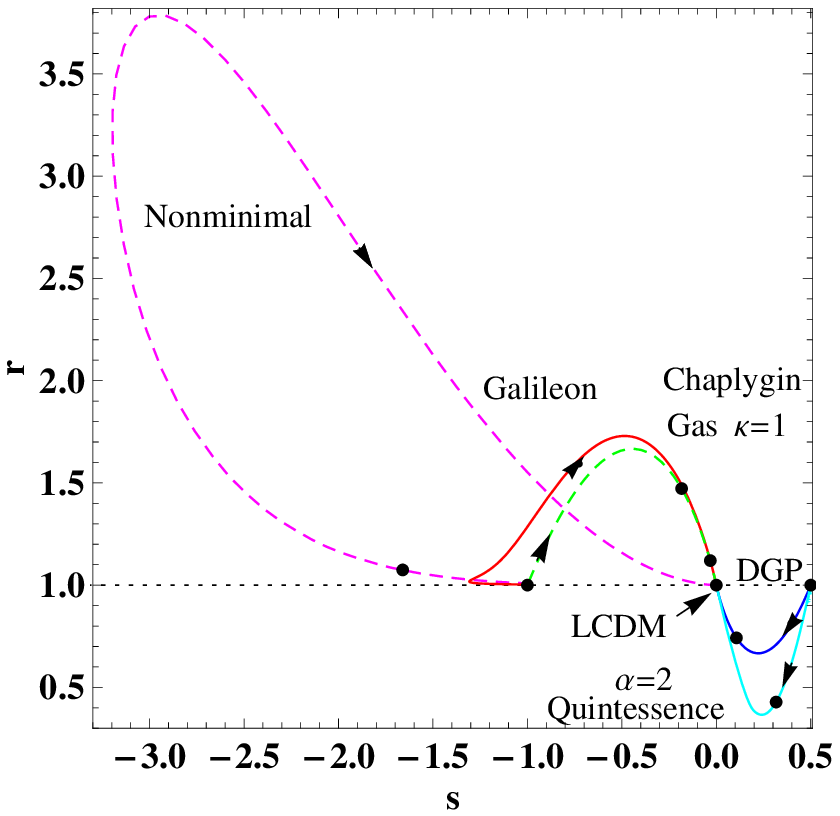} &
        \epsfxsize=3.2in
        \epsffile{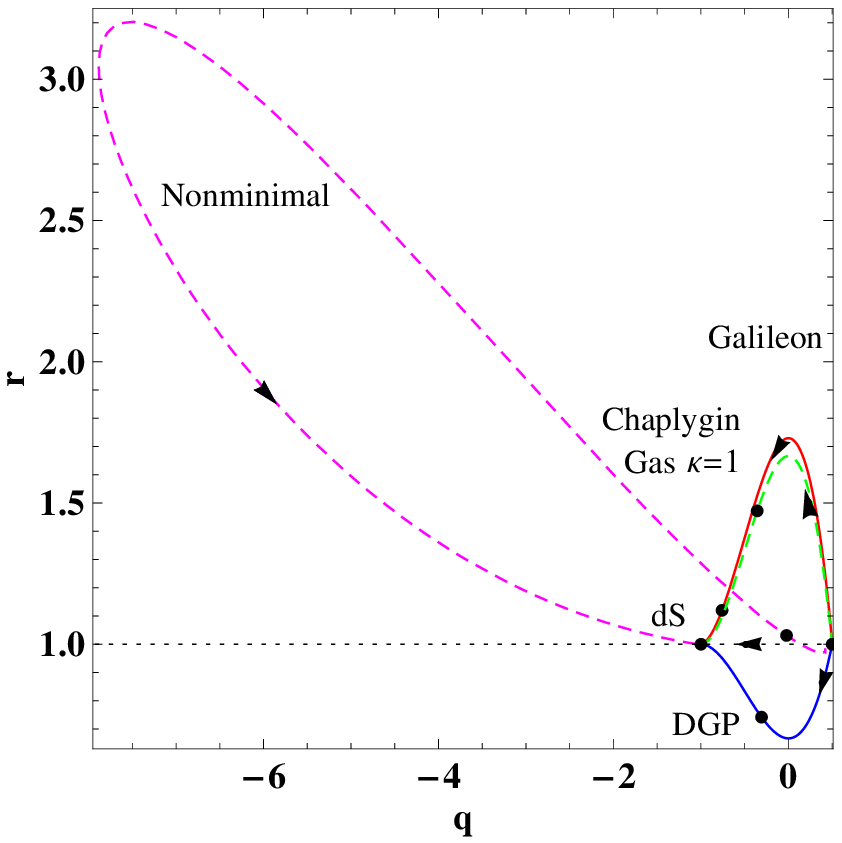} \\ [0.20cm]
\mbox{\bf (a)} & \mbox{\bf (b)}
\end{array}$
\end{center}
\caption{(color online). The left panel (a) shows the time evolution
of the statefinder pair $\lbrace r,s \rbrace$ for Galileon (red),
 Nonminimal (magenta), DGP (blue), quintessence (Cyan) and Chaplygin gas
  (green) models. All models converge to the fixed point ($r=1, s=0$)
  which corresponds to $\Lambda$CDM.The right panel (b) shows the
  time evolution of the statefinder pair $\lbrace r,q\rbrace$ for
   Galileon (red), nonminimal (magenta), DGP (blue) and chaplygin
   gas (green) models.  All models diverge at the same point ($r=1,q=0.5$)
   which corresponds to a matter dominated universe (SCDM),
   and converge to the same point ($r=1,q=-1$) which corresponds
   to the de Sitter expansion(dS).The dark dots on the curves show
   current values $\lbrace r_0,s_0 \rbrace$ (left) and
    $\lbrace r_0,q_0 \rbrace$ (right) for different
     dark energy models. In all models, $\omm = 0.3$
      at the current epoch.}
  \label{stat2}
\end{figure*}
\section{Comparison with the competing models }
There are several popular models of dark energy which can comply
well with popular observational data  sets. The list of models is
rather large, we shall restrict to commonly used models: $\Lambda
CDM$, Quintessence with tracking behavior, Chaplygin gas, DGP and
recently discussed galileon model which generically belongs
to the class of non-minimal coupled models.\\
One of the methods to distinguish between various models at the
background level is provided by statefinder analysis. The Hubble
parameter $H$ and the deceleration parameter $q$ constructed from
the first and the second derivatives of the scale factor are often
used to characterize a cosmological model. The statefinder method
relies on the use of parameters constructed from third derivatives
(and higher) of the scale factor\cite{statefinder,maryam}. For
instance, in the lowest order of this hierarchy, we have
\begin{eqnarray}
r=\frac{\dddot{a}}{aH^3},~~~~s=\frac{r-1}{3(q-1/2)}=1+w-\frac{1}{3}\frac{\dot{w}}{wH},
\end{eqnarray}
which shows that $(r,s)=(1,0)$ $\&$ $(1,1)$ for $\Lambda CDM$ and
matter dominated universe respectively. In what follows, we shall
briefly describe the aforementioned dark energy models followed by
their statefinder analysis.

\section{DARK ENERGY MODELS}
\label{sec2}
\begin{itemize}
\item {\bf Galileon  model}\cite{Gannouji:2010au}.\\

Galileon is a massless scalar field, its Lagrangian apart from the
kinetic term contains non-linear derivative terms responsible for
switching off the modification to gravity locally. The theory is
free from Ostrogradki ghosts despite the higher order derivative
terms in the Lagrangian. The galileon action is of the form, \be
\mathcal{S}=\int{\textit{\rd}^4 x}\sqrt{-g}
\left(\frac{R}{2}+c_\textit{i}~L^{(\textit{i})}\right)+\mathcal{S}_m[\psi_m,e^{2\beta\pi}g_{\mu\nu}]
\label{action} \ee
where $\{c_i\}$ are constants, $\beta$ is coupling constant of field
with matter and  $L_i^{'s}$ \cite{Gannouji:2010au} are Galileon
Lagrangians. $L_1$ is linear in field and is often omitted assuming
$c_1=0$, $L_2$ represents the standard kinetic terms,
$L_3=(\partial_\mu \pi)^2 \Box \pi$ is the famous Vainshtein term in
which   three $\pi's$ participate; $L_4$ and $L_5$ are higher order
Lagrangians. At least $L_4$ is needed to obtain dark energy solution
in the model\cite{Gannouji:2010au}. In this case, the evolution
equations in a spatially flat background  have the
form\cite{Gannouji:2010au}
\begin{eqnarray}
\label{eq:Fried1}
3H^2&=&\rho_m+\frac{c_2}{2}\dot{\pi}^2-3c_3 H\dot{\pi}^3+\frac{45}{2}c_4 H^2\dot{\pi}^4,\\
\label{eq:Fried2}
2\dot{H}+3H^2&=&-\frac{c_2}{2}\dot{\pi}^2-c_3\dot{\pi}^2\ddot{\pi}+\frac{3}{2}c_4\dot{\pi}^3\left(3H^2\dot{\pi}+2\dot{H}\dot{\pi}+8H\ddot{\pi}\right),
\end{eqnarray}
\begin{eqnarray}
\label{eq:KG}
\beta\rho_m&=&-c_2\left(3H\dot{\pi}+\ddot{\pi}\right)+3c_3\dot{\pi}\left(3H^2\dot{\pi}+\dot{H}\dot{\pi}
+2H\ddot{\pi}\right)
-18c_4H\dot{\pi}^2\left(3H^2\dot{\pi}+2\dot{H}\dot{\pi}+3H\ddot{\pi}\right)
\end{eqnarray}

 The energy density, pressure for the scalar field $\pi$ and the Friedmann equation are,
\begin{eqnarray}
\label{eq:rhopi}
\rho_{\pi}&=&\frac{c_2}{2}\dot{\pi}^2-3c_3 H\dot{\pi}^3+\frac{45}{2}c_4 H^2\dot{\pi}^4,\\
p_{\pi}&=&\frac{c_2}{2}\dot{\pi}^2+c_3\dot{\pi}^2\ddot{\pi}-\frac{3}{2}c_4\dot{\pi}^3\left(3H^2\dot{\pi}+2\dot{H}\dot{\pi}
+8H\ddot{\pi}\right)\nonumber\\
H^2 &=& \frac{8\pi G}{3}\left\lbrack \rho_{0m}(1+z)^3 +\rho_{\pi}
\right\rbrack;~~~w_\pi=\frac{p_\pi}{\rho_\pi}
\end{eqnarray}

As demonstrated in Ref.\cite{Gannouji:2010au}, there is a region in
phase space which necessarily contains a stable de Sitter solution.
In Fig.\ref{wpi}, we have plotted equation of state parameter
$w_{\pi}$ (with favorable  initial conditions) which shows that the
latter ultimately approaches its de Sitter value.
\item {\bf Dark energy with a constant equation of state}\cite{review1}.

 The Hubble
parameter is given by,
\begin{equation}
\label{eq:lambda}
H(z) = {H_0} \left\lbrack \omm(1+z)^3 +
\Omega_{\rm DE}(1+z)^{3(1+w)}\right\rbrack^{1/2}~
\end{equation}
where ($\Lambda$CDM) corresponds to $w=-1$ and for Quiessence \cite{statefinder} $w = {\rm constant} \neq -1$.

\item {\bf Non-minimally coupled scalar field}\cite{Polarski}.
 The energy density and pressure in this case are given by,
\begin{eqnarray}
\label{eq:rhophi}
\rho_{\phi}&=&\frac{1}{2}{\dot {\phi}}^{2}+V(\phi)+3\xi( H \dot {\phi} B'(\phi)+H^{2} B(\phi)),\\
p_{\phi}&=&\frac{1}{2}{\dot {\phi}}^{2}-V(\phi)-\xi\left(2 H \dot
{\phi} B'(\phi)+ \dot {\phi}^{2} B''(\phi)+\ddot {\phi}
B'(\phi)+(2\dot{H}+3H^{2}) B(\phi)\right).
\end{eqnarray}
The Hubble parameter for this model has the form,
\be
\label{eq:nonminimal} H^2 = \frac{8\pi G}{3}\left\lbrack
\rho_{0m}(1+z)^3 +\rho_{\phi} \right\rbrack;~~~~w_{\phi} =
\frac{p_{\phi}}{\rho_{\phi}}
 \ee
\item {\bf  Quintessence}\cite{review1,review2}.

In this case, there is a variety of models which can lead to desired
cosmic evolution with a carefully chosen field potential. All the
quintessence models can be classified into two groups: Thawing and
Freezing or tracker models. In case of thawing, cosmic evolution
crucially depends upon initial conditions where as the tracker
dynamics is free from the choice of initial conditions and we shall
focus on the latter. Bearing that in mind, we have the expression
for Hubble parameter,
\begin{eqnarray}
\label{eq:quint} H^2&=&\frac{8\pi G}{3}\left\lbrack \rho_{0m}(1+z)^3
+\rho_{\phi} \right\rbrack;~~~V(\phi) \propto \phi^{- \alpha},
\alpha \geq1
\end{eqnarray}

\item {\bf Chaplygin gas \cite{chap1}}.
\begin{equation}
\label{eq:chap} H(z) = {H_0} \left\lbrack \omm(1+z)^3 +
\frac{\omm}{\kappa}\sqrt{\frac{A}{B}
+(1+z)^6}\right\rbrack^{1/2}~;~~ \kappa = \rho_{0m}/\sqrt{B} ,~~A =
B \left\lbrace \kappa^2 \left( \frac{1-\omm}{\omm} \right)^2 - 1
\right\rbrace \,\,.
\end{equation}

\item  {\bf The
Dvali, Gabadadze , Porrati (DGP) model} \citep{dgp}:
\begin{equation}
\label{eq:DGP}
\frac{H(z)}{H_0} = \left[ \left(\frac{1-\omm}{2}\right)+\sqrt{\omm (1+z)^3+
\left(\frac{1-\omm}{2}\right)^2} \right]\,\,.
\end{equation}
\end{itemize}
In Fig.\ref{stat0}, we have shown the time evolution of the
statefinder pairs $\lbrace r,s \rbrace$ and $\lbrace r,q \rbrace$
for non-minimally coupled system. The model converge to the fixed
point ($r=1, s=0$) which corresponds to $\Lambda$CDM. The right
panel (b) shows the time evolution of the statefinder pair $\lbrace
r,q\rbrace$, the model diverge at the point ($r=1,q=0.5$) which
corresponds to a matter dominated universe (SCDM), and converge to
the point ($r=1,q=-1$) which corresponds to the de Sitter
expansion(dS).\\
For comparision with other models, we have displayed our model along
with other competing models in Figs.(\ref{stat1})$\&$ (\ref{stat2}).
Since, to the best of our knowledge, statefinder analysis for
galileon is not done elsewhere, we have in Fig.(\ref{stat1})
displayed $(r,s)$ trajectories for non-minimally coupled scalar
field system and galileon field which is generically non-minimally
coupled system. Fig.\ref{stat2} displays statefinder trajectories
for non-minimally coupled scalar field along with other popular
models of dark energy including the galileon field. The figure
 shows
that
 all models diverge from the same point ($r=1,q=0.5$)
  which corresponds to a matter dominated universe (SCDM) and converge to the  point ($r=1,q=-1$)
  which represents de Sitter universe(dS). Fig.\ref{stat1} shows
  how clearly the two non-minimally coupled systems get
  distinguished on the (r,s) $\&$ (r,q) planes. In case of galileon,
  we focussed
  on the sector of the model which contains stable de
  Sitter fixed points. Fig.\ref{stat2}
allows to distinguish these two models from other popular models
  of dark energy. It may be noted that we have re-scaled appropriately the statefinder variables
  of the model under consideration in order to strike comparison with other models. It is interesting
  to note  that the behavior of statefinder
  pairs for galileon and Non-minimally  coupled scalar field
systems resembles with that of the braneworld models studied in
Ref.\cite{uv}.



\section{Conclusion}
In this paper, we have revisited cosmological dynamics of
non-minimally coupled scalar field system. We presented detailed
investigation of dynamics of the underlying system  in case of
$F(\phi)R=(1-\xi B(\phi))R$ coupling with $B(\phi)=\phi^N$ and $
V(\phi)=V_0\phi^n$ ($N\geq 2$) using a convenient set of autonomous
variables. We studied asymptotic regimes of solutions in the model.
In case of the vacuum solution, we found a very interesting solution
for which $\dot{H}=0$ ($w_{eff}=-1$) and the scalar field $\phi$
increasing exponentially giving rise to exponentially decreasing
$G_{N}$ $-$ effective Newtonian gravitational constant. Such a
solution does not qualify for de
Sitter for which $G_N$ should be held constant. \\
The autonomous variables used in this paper though convenient in
general but miss certain important features of the dynamics. The
description fails to capture the de Sitter solution as the
combination of variables $4x+z^2$ appearing in the denominator of
the autonomous system vanishes identically in this case. The
investigation of this solution took us back to the original variable
in the evolution equations. We found that in case of de Sitter
solution, $G_N>0$ provided that $n<2N$. On the other hand, our
numerical investigations showed that the solution under
consideration is stable only for $n\geq 2N+1$ (we checked for lower
values of $N \ge 2$). For initial conditions of matter dominated
universe, the system enters the phase of acceleration consistent
with observation at present followed by a brief phantom phase
thereafter which  continues  till de Sitter is reached, see
Fig.\ref{gw}(b) (which is an attractive focus of the dynamics, see
Fig.\ref{vds}(b)). During the phantom phase $G_N$ changes sign from
positive to negative thereby making the universe ghost dominated in
future, see Fig.\ref{gw}. It is possible to set parameters in the
model such that phantom phase occurs at present epoch corresponding
to $G_N>0$ compilable with observed values of the equation of state
and fractional density parameter pushing the ghost dominated phase
to future which no body has yet seen. Incidentally, similar features
of equation of state  are shared by the braneworld model discussed
in Ref.\cite{yv}.\\
An important remark about the ghost dominated universe with $G_N<0$
is in order. {\it It was demonstrated by Starobinsky in 1981 that
any attempt of crossing over to the regime with $G_N<0$ makes the
Friedmann universe unstable at the transition point. Any tiny
perturbation over homogeneity and isotropy grows large at the point
of transition. Taking into account the quantum effects in the scalar
sector does not affect this result\footnote{ Though the higher order
curvature corrections to Einstein-Hilbert action motivated by
Starobinsky in different context could help to avoid the catastrophe
and land the theory ghost free. }. The ghost dominated universe with
$G_N<0$, if exists, will be separated from our real universe by a
region of large curvature of the order of Planck value. Thus it is
impossible to cross the point where the effective gravitational
constant changes sign due to the formation of generic anisotropic
curvature singularity at this point}\cite{STN}.

The de Sitter solution for the case with $N=2$ also shares the
aforementioned features. In other cases, we have shown that the
solutions obtained earlier are continued for values of $N>2$. We
have investigated in detail the asymptotic regimes of these
solutions in all cases including the one corresponding to $N=2$

We have shown that the non-minimally coupled scalar field system can
account for late time cosmic acceleration. We should, however,
emphasize that dark energy , in certain sense, in this scenario
appears as a transient phenomenon which involves extra fine tuning.
{ The transient phase is followed by a stable de Sitter point which
lies in the regime of negative value of effective gravitational
constant and can not be reached from the Universe we live in, a
curvature singularity separates us from the de Sitter fixed point.
In view of the phase space evolution, all the trajectories, at the
present epoch, nearly converge to the one with equation of state
parameter around $-1$. The small change of initial conditions
slightly changes the present value of equation of state parameter of
dark energy. Tuning the initial conditions we can achieve phantom or
non-phantom dark energy consistent with observations on late time
cosmic acceleration.}


Last but not least, we used statefinder analysis to compare the
model with other rival models of dark energy, in particular, with
galileon model which generically a non-minimally coupled system. The
model under consideration is clearly distinguished on the $(r,s)$
and $(r,q)$ planes from other competing models of dark energy. We
focussed our attention on sectors which contain stable de Sitter
solution in case of the model under consideration as well as the
galileon modified theory of gravity.

A final remark about the ghost dominated evolution in the model is
in order. In the framework of the simple set up of non-minimal
scheme discussed here, there exists no consistent de Sitter solution
such that $G_N$ remains positive throughout the evolution. It would
really be interesting to explore generic functional forms of the
coupling function and the field potential to check for a well
behaved de Sitter solution.
\section*{Acknowledgments}
We are indebted to A. starobinsky for his kind comment on the
impossibility of crossing over to ghost dominated universe. We thank
V. Sahni, E. Saridakis and S. Yu. Svernov for useful comments. This
work was supported in part by the FTsP ``Nauchnie i
nauchno-pedagogicheskie kadry innovatsionnoy Rossii'' for the years
2009-2013". MSA acknowledge the financial support provided by the
DST, Government of India, through the research project N0.
SR/S2/HEP-002/2008. He is thankful to Amna Ali and Wali Hossain for
fruitful discussions. MS thanks S. Nojiri and K. Bamba for useful
discussions. M.S. is supported by the JSPS Invitation Fellowship for
Research in Japan (Long-Term) \# L12422 and also by the Department
of Science and Technology, India. MS thanks Kobayashi-Maskawa
Institute for the Origin of Particles and the Universe for
hospitality. AT is supported by RFBR grant no 11-02-00643 of the
Russian Ministry of Science and Technology.

\section{Appendix}
\label{subcase}
\subsection{The case of $b=1(N=2)$.}
As mentioned before, this case was considered earlier in other
references. However, discussion of asymptotic regimes  was missing
and important issues related to the structure of de Sitter solution
were not discussed. Thus we have retained this subsection in the
appendix which is also useful for comparison of our results with
$b=1(N=2)$ case.

 As mentioned above, there exists a simple relation between $x$ and $z$ for $N=2$.
 Indeed we have,
\begin{eqnarray}
\label{eq:b1} x&=&\frac{\dot{\phi^2}}{H^2(1-6\xi B(\phi))}
=\frac{\dot{\phi^2}}{H^2(1-6\xi B(\phi))}\frac{{(6\xi
B'(\phi))}^2}{{(6\xi B'(\phi))}^2} \frac{\phi(1-6\xi
B(\phi))}{\phi(1-6\xi B(\phi))}=\frac{z^2}{72\xi^2  A}
\end{eqnarray}

 Now we substitute (\ref{eq:b1}) and $b=1$ in  (\ref{eq:capxy}),
 \begin{eqnarray}
\label{eq:b3}
\Omega&=&1-\frac{z^2}{72{\xi}^2 A}-y-z,\nonumber\\
X&=&-\frac{z}{2}-\frac{1}{1+18\xi^2 A}\left (\frac{z^2}{12}(6\xi -1)+y\xi (12\xi A+c) \nonumber \right. \\
&&\left.+3\xi^2 A(1-\frac{z^2}{72\xi^2 A}-y-z)(1-3\omega)\right ),\nonumber\\
Y&=&\frac{1}{1+18\xi^2  A}\left (\frac{z^2}{12\xi^2 A}(6\xi -1)+6y\nonumber \right. \\
&&\left.(2-3 c\xi )+3(1-\frac{z^2}{72\xi^2  A}-y-z)(1-3\omega)\right
).
\end{eqnarray}
Excluding from (\ref{eq:autonomous4}) $x$, $\Omega$, $X$, $Y$ using
(\ref{eq:b3}), we obtain following system for $b=1$ ($N=2$)
\begin{eqnarray}
\label{eq:b4}
y'&=&\frac{yz}{6\xi}\frac{c}{A}-2y\Big{(}\frac{1}{6(1+18\xi^2  A)}\left (\frac{z^2}{12\xi^2 A}(6\xi -1)+6y
 (2-3 c\xi ) +3(1-\frac{z^2}{72\xi^2  A}-y-z)(1-3\omega)\right )-2\Big{)}
+yz, \nonumber \\
z'&=&\left( -3z-\frac{6}{1+18\xi^2  A}\Big{(}\frac{z^2}{12}(6\xi
-1)+y\xi (12\xi A+c)+3\xi^2 A(1-\frac{z^2}{72\xi^2
A}-y-z)(1-3\omega) \Big{)}\right)\nonumber\\
&& +\frac{z^2}{6\xi A}-z\left( \frac{1}{6(1+18\xi^2
A)}\Big{(}\frac{z^2}{12\xi^2 A}(6\xi -1)+6y (2-3 c\xi )
+3(1-\frac{z^2}{72\xi^2  A}-y-z)(1-3\omega) \Big{)}-2 \right)
+z^2,\nonumber\\
 A'&=&\frac{z}{3\xi}+Az.
\end{eqnarray}
  We enumerate the stationary points and corresponding solutions of
  (\ref{eq:b4}),\\

\textbf{1. $de$-Sitter solution}
\begin{eqnarray}
\label{eq:b5}
y&=&1, z=0, A=-\frac{c}{12\xi}, \Omega=0,\nonumber\\
Y&=&12, \beta=0,\nonumber\\
a(t)&=&a_0 e^{H_0(t-t_0)},\nonumber\\
\phi(t)&=&\phi_0,
\end{eqnarray}
where $\phi_0=\pm\sqrt{\frac{n}{6\xi (n-4)}}$,
$H_0=\pm\sqrt{-\frac{V_0 n {\phi_0}^{n-2}}{12\xi }}$ are obtained
from the system (\ref{eq:Friedphi})- (\ref{eq:KGphi}) for
$\dot{H_0}=\dot{\phi_0}=\ddot{\phi_0}=\rho=0$. This is de Sitter
solution which exists also for $N>2$. We have not solved the
autonomous system  in this case
 as the combination $4x+z^2$ appearing in the denominator  vanishes in case of de Sitter solution
 with
 coordinates $x=0$, $y=1$, $z=0$, $A=-\frac{n}{12\xi}$,
 $\Omega=0$.\\

\textbf{2.}
\begin{eqnarray}
\label{eq:b6}
y&=&0, z=12 \xi+2 \sqrt{36 {\xi}^2-6 \xi} ,\nonumber\\
A&=&-\frac{1}{3\xi}, \Omega=0,\nonumber\\
Y&=&6(12 \xi-1+2 \sqrt{6  \xi(6 \xi-1)} ), \nonumber\\
\beta&=&-6 \xi-\sqrt{6  \xi(6  \xi-1)},\nonumber\\
a(t)&=&a_0{|t-t_0|}^{\frac{1}{3-12 \xi-2 \sqrt{6 \xi(6 \xi-1)} }},\nonumber\\
\phi(t)&=&{\phi}_0{|t-t_0|}^{-\frac{6 \xi+\sqrt{6 \xi(6
\xi-1)}}{3-12  \xi-2 \sqrt{6  \xi(6 \xi-1)}}},
\end{eqnarray}
\\

\textbf{3.}
\begin{eqnarray}
\label{eq:b7}
y&=&0, z=12 \xi-2 \sqrt{36{\xi}^2-6 \xi} ,\nonumber\\
A&=&-\frac{1}{3\xi}, \Omega=0,\nonumber\\
Y&=&6(12 \xi-1-2 \sqrt{6 \xi(6 \xi-1)} ),\nonumber\\
\beta&=&-6 \xi+\sqrt{6  \xi(6  \xi-1)},\nonumber\\
a(t)&=&a_0{|t-t_0|}^{\frac{1}{3-12 \xi+2 \sqrt{6 \xi(6 \xi-1)} }},\nonumber\\
\phi(t)&=&{\phi}_0{|t-t_0|}^{\frac{-6 \xi+\sqrt{6 \xi(6
\xi-1)}}{3-12 \xi+2 \sqrt{6 \xi(6  \xi-1)}}},
\end{eqnarray}
Solutions corresponding to stationary points 2, 3 were found in
\cite{Dunsby1}, where the cosmological model with non-minimal
coupling $F(\phi)=\xi\phi^2$ and the effective potential
$V(\phi)=\lambda\phi^n$ (without Einstein term in the Lagrangian)
was investigated.
\\

\textbf{4.}
\begin{eqnarray}
\label{eq:b7}
y&=&0, z=-\frac{4 \xi(1-3\omega)}{4 \xi+\omega-1}, A=-\frac{1}{3\xi},\nonumber\\
\Omega&=&1+\frac{2  \xi{(1-3\omega)}^2}{3{(4 \xi+\omega-1)}^2}+\frac{4 \xi(1-3\omega)}{4 \xi+\omega-1},\nonumber\\
Y&=&\frac{3(1-3\omega)(\omega-1)}{4 \xi+\omega-1}, \beta=\frac{2 \xi(1-3\omega)}{4  \xi+\omega-1},\nonumber\\
a(t)&=&a_0{|t-t_0|}^{\frac{2(4 \xi+\omega-1)}{3\omega^2+16 \xi-3 }},\nonumber\\
\phi(t)&=&{\phi}_0{|t-t_0|}^{\frac{4 \xi(1-3\omega)}{3\omega^2+16
\xi-3 }},
\end{eqnarray}
This solution also was found in \cite{Dunsby1}.
\\

\textbf{5.}
\begin{eqnarray}
\label{eq:b8}
y&=&-\frac{(6 \xi\omega(c-2)-2 \xi(c+6)+3-3{\omega}^2)}{2 c^2 \xi},\nonumber\\
z&=&\frac{6(1+\omega)}{c},A=-\frac{1}{3\xi},\nonumber\\
\Omega&=&\frac{3(1-\xi(c+2))\omega+c^2 \xi+3-7 c \xi-6 \xi}{c^2  \xi},\nonumber\\
Y&=&\frac{3(6+c-3\omega(c-2))}{c}, \beta=-\frac{3(1+\omega)}{c},\nonumber\\
a(t)&=&a_0{|t-t_0|}^{\frac{2n}{3(n-2)(1+\omega)}},\nonumber\\
\phi(t)&=&{\phi}_0{|t-t_0|}^{\frac{2}{2-n}},\nonumber\\
\rho(t)&=&{\rho}_0{|t-t_0|}^{\frac{2n}{2-n}}.
\end{eqnarray}
We note that the analogous solution exists in the case $N>2$ (see
(\ref{eq:rho11})). It also was obtained in \cite{Dunsby1}.
  For $n=2$ the power index of functions $a(t)$ and $\phi(t)$ becomes infinity and power-law
  takes the exponential form. In this case the coordinates of the point 5
   are $y=\frac{16\xi-3+3\omega^2}{8\xi}, z=3(\omega+1), A=-\frac{1}{3\xi}, \Omega=\frac{3(1-4\xi)\omega-16\xi+3}{4\xi}$
    and also, $Y=12$, $\beta=-\frac{3(1+\omega)}{2}$. This means that, $\frac{\dot{H}}{H^2}=0$, $H=H_0=const$,
    $\frac{\dot{\phi}}{\phi}=\beta H_0$ giving rise to the following

\begin{eqnarray}
\label{eq:b9}
a(t)&=&a_0 e^{H_0(t-t_0)},\nonumber\\
\phi(t)&=&\phi_0 e^{\beta H_0(t-t_0)}=\phi_0 e^{-\frac{3 H_0(1+\omega)(t-t_0)}{2}},\nonumber\\
\rho(t)&=&{\rho}_0 e^{-3 H_0(1+\omega)(t-t_0)}.
\end{eqnarray}
 In order to find the constants $H_0$ and $\rho_0$, we substitute (\ref{eq:b9}) in definition (\ref{eq:omega}) of
 coordinates $y$, $\Omega$ taking into account that $n=2$, $y=\frac{16\xi-3+3\omega^2}{8\xi}$
 and $\Omega=\frac{3(1-4\xi)\omega-16\xi+3}{4\xi}$,
\begin{eqnarray}
\label{eq:b10}
y&=&\frac{2V_0\phi^n}{H^2(1-6\xi\phi^2)}=\frac{2V_0{\phi_0}^2 e^{-3
H_0(1+\omega)(t-t_0)}}{{H_0}^2\left( 1-6\xi{\phi_0}^2 e^{-3
H_0(1+\omega)(t-t_0)}\right) }\rightarrow -\frac{V_0}{3{H_0}^2 \xi}\nonumber\\
\Omega&=&\frac{2\rho}{H^2(1-6\xi\phi^2)}= \frac{2\rho_0 e^{-3
H_0(1+\omega)(t-t_0)}}{{H_0}^2\left( 1-6\xi{\phi_0}^2 e^{-3
H_0(1+\omega)(t-t_0)}\right) }\rightarrow -\frac{\rho_0}{3{H_0}^2
{\phi_0}^2\xi}
\end{eqnarray}
\\
\text{for $t\rightarrow t_0$ when $H_0>0$(or for $t\rightarrow
\infty $ when $H_0<0$)}.
\\
Therefore, \be H_0^2=-\frac{V_0}{3\xi y}=-\frac{8
V_0}{3(16\xi-3+3\omega^2)},\nonumber \ee
 where $\xi<\frac{3(1-\omega^2)}{16}$, $V_0>0$ or $\xi>\frac{3(1-\omega^2)}{16}$, $V_0<0$
 and
\be \rho_0=-3\Omega{H_0}^2 {\phi_0}^2\xi=\frac{2V_0
{\phi_0}^2(3(1-4\xi)\omega-16\xi+3)}{16\xi-3+3\omega^2}\nonumber \ee

\textbf{6.}
\begin{eqnarray}
\label{eq:b11}
y&=&-\frac{(6 c^2 {\xi}^2-c^2 \xi+8 c \xi-48{\xi}^2 c-120 {\xi}^2+56  \xi-6)}{6{( \xi(2+c)-1)}^2}, \nonumber\\
z&=&\frac{2  \xi(c-4)}{ \xi(2+c)-1}, A=-\frac{1}{3\xi}, \Omega=0,\nonumber\\
Y&=&\frac{3(10 c \xi-c^2 \xi-4)}{\xi(2+c)-1}, \beta=-\frac{\xi(c-4)}{\xi(2+c)-1},\nonumber\\
a(t)&=&a_0{|t-t_0|}^{\frac{2 (\xi(2+n)-1)}{\xi(n-2)(n-4)}},\nonumber\\
\phi(t)&=&{\phi}_0{|t-t_0|}^{\frac{2}{2-n}}
\end{eqnarray}
This solution in the limit $\xi\rightarrow\infty$ gives
(\ref{eq:phi1})\cite{Dunsby1}.

    For $n=2$ and $n=4$ power indexes of functions $a(t)$ and $\phi(t)$ become infinity and
    power-law solutions transform into exponential ones. We find
    that coordinates of point 6 in these cases are, $y=\frac{96\xi^2-34\xi+3}{3{(4\xi-1)}^2}, z=-\frac{4\xi}{4\xi-1}, A=-\frac{1}{3\xi}, \Omega=0$ and $y=1, z=0, A=-\frac{1}{3\xi}, \Omega=0$ accordingly for $n=2$ and $n=4$. Then $Y=12$, $\beta=\frac{2\xi}{4\xi-1}$ -- for $n=2$ and $Y=12$, $\beta=0$ -- for $n=4$. For $n=2$ we obtain $\frac{\dot{H}}{H^2}=0$, $H=H_0=const$ and
\begin{eqnarray}
\label{eq:b12}
a(t)&=&a_0 e^{H_0(t-t_0)},\nonumber\\
\phi(t)&=&\phi_0 e^{\beta H_0(t-t_0)}=\phi_0 e^{\frac{2
H_0\xi(t-t_0)}{4\xi-1}}.
\end{eqnarray}
Using the definition of the coordinate $y$ from (\ref{eq:omega})
analogous to previous point we find the constant $H_0$ ($n=2$,
$y=\frac{96\xi^2-34\xi+3}{3{(4\xi-1)}^2}$) and
 hence ${H_0}^2=-\frac{V_0}{3\xi y}=-\frac{V_0{(4\xi-1)}^2}{\xi(96\xi^2-34\xi+3)}$, where $\xi\in(-\infty,0)\bigcup(\frac{1}{6},\frac{9}{48})$, $V_0>0$ or $\xi\in(0,\frac{1}{6})\bigcup(\frac{9}{48})$, $V_0<0$.
 For $n=4$ we get $\frac{\dot{\phi}}{\phi}=0$ and
\begin{eqnarray}
\label{eq:b13}
a(t)&=&a_0 e^{H_0(t-t_0)}=a_0,\nonumber\\
\phi(t)&=&\phi_0=0,
\end{eqnarray}
where values $H_0=0$, $\phi_0=0$ were found by substitution
$a(t)=a_0 e^{H_0(t-t_0)}$, $\phi(t)=\phi_0$, $n=4$ in the system of
equations (\ref{eq:Friedphi})-(\ref{eq:KGphi}).


\begin{thebibliography}{99}
\bibitem{BD} C. Brans and R. Dicke, Phys. Rev.  {\bf 124}, 925 (1961).
\bibitem{BS} F.Bezrukov and M.Shaposhnikov, Phys.Lett. {\bf B639}, 703 (2008).
\bibitem{PR1} S. Perlmutter {\it el al}, Astrophysics, J. {\bf 157},
565(1999).
\bibitem{PR2} A. Reiss {\it et al}, Astrophysics J. {\bf 117},
707(1999).

\bibitem{review1}
V.~Sahni and A.~A.~Starobinsky, Int.\ J.\ Mod.\ Phys.\ D \textbf{9},
373 (2000).
\bibitem{vpaddy} V. Sahni and A. Starobinsky,
     Int.J.Mod.Phys.D {\bf 15}, 2105(2006)[astro-ph/0610026]; T. Padmanabhan,
     astro-ph/0603114;
 P.~J.~E.~Peebles and B.~Ratra, Rev.\ Mod.\ Phys.\ {} \textbf{75},
559 (2003); L. Perivolaropoulos, astro-ph/0601014; N. Straumann,
arXiv:gr-qc/0311083;  J. Frieman, arXiv:0904.1832; M. Sami, Lect.
Notes Phys.{\bf 72}, 219(2007); M. Sami,  arXiv:0901.0756 ; K.
Bamba, S. Capozziello, S. Nojiri and S. D. Odintsov,
arXiv:1205.3421; S. Tsujikawa, arXiv:1004.1493.

\bibitem{review2}E.~J.~Copeland, M.~Sami and
S.~Tsujikawa, Int. J. Mod. Phys., {D15} ,
1753(2006)[hep-th/0603057].
\bibitem{review3} E. V. Linder,  Rep. Prog. Phys. {\bf 71} (2008)
056901.
\bibitem{review3C} Robert R. Caldwell and Marc
Kamionkowski,arXiv:0903.0866.
\bibitem{review3d} A. Silvestri and Mark Trodden, arXiv:0904.0024.
\bibitem{review4}J. Frieman, M. Turner and D.
Huterer, arXiv:0803.0982.
\bibitem{review5} M. Sami, Curr. Sci. 97,887(2009)[arXiv:0905.2284]
\bibitem{Polarski} R.Gannouji, D.Polarski, A.Ranquet and A.Starobinsky, JCAP 0609:016 (2006).
\bibitem{Tsujikawa} L.Amendola, R.Gannouji, D.Polarski, S.Tsujikawa, Phys. Rev. {\bf D75},
083504 (2007).
\bibitem{Dunsby} S.Carloni, A.Troisi, P.Dunsby, Gen.Rel.Grav.{\bf 41}, 1757 (2009).
\bibitem{Dunsby1} S.Carloni, J.Leach, S.Capozziello, P.Dunsby. Class. Quant. Grav. 25, 035008 (2008).
\bibitem{mark}M. Szydlowski and O. Hrycyna, {JCAP} {0901}, {039}
{(2009)}[{arXiv:0811.1493v2}].
\bibitem{sergei}
 E. Elizalde, S. Nojiri, S. D. Odintsov, D. Sez-Gmez and V. Faraoni,
Phys.Rev.D77:106005,2008[arXiv:0803.1311].
\bibitem{Kuusk} L.Jarv, P.Kuusk, M.Saal, Phys. Rev. {\bf D76}: 103506 (2008),
Phys. Rev. {\bf D78}: 083530 (2008), Phys. Rev. {\bf D81}: 104007
(2010), Phys. Lett. {\bf A694}, 1 (2010); Gaveshna Gupta, Emmanuel
N. Saridakis, and Anjan A. Sen,
  Phys.\ Rev.\  D {\bf 79} (2009) 123013;
  Y.~-F.~Cai, E.~N.~Saridakis, M.~R.~Setare and J.~-Q.~Xia,
  Phys.\ Rept.\  {\bf 493}, 1 (2010)
  [arXiv:0909.2776 [hep-th]]; E.~Elizalde, E.O.~Pozdeeva and S.Yu.~Vernov,
De Sitter Universe in Non-local Gravity,
  Phys. Rev. D {\bf 85} (2012) 044002, arXiv:1110.5806

\bibitem{Kamen} A.Yu.Kamenshchik, I.M.Khalatnikov, A.V.Toporensky, Int. J. Mod. Phys. D6 (1997), 649.
\bibitem{Gannouji:2010au}
  Amna Ali, R.~Gannouji and M.~Sami,
  Phys.\ Rev.\  D {\bf 82} (2010) 103015;
  R.~Gannouji and M.~Sami,
  Phys.\ Rev.\  D {\bf 82} (2010) 024011;

\bibitem{statefinder}
V. Sahni, T.~D. Saini, A.~A.~Starobinsky and U.~Alam, \jetpl
{\bf 77}, 201 (2003); 
U. Alam, V. Sahni, T.~D. Saini and A.~A.~Starobinsky, \mn
{\bf 344}, 1057 (2003). 
\bibitem{maryam}
M.~Arabsalmani, V.~Sahni, Phys. Rev. D {\bf 83}, 043501
(2011).
\bibitem{sf-phantom}
{B.\,R.~Chang, H.\,Y.~Liu, L.\,X.~Xu, C.\,W.~Zhang and
Y.\,L.~Ping,} {Statefinder parameters for interacting phantom energy
with dark matter,}{JCAP} {0701}:{016} {(2007)} [{astro-ph/0612616}].
%
\bibitem{Kam-sf-in-Chap}
{V.~Gorini, A.~Kamenshchik and U.~Moschella,} {Can the
Chaplygin gas be a plausible model for dark
energy?} {Phys.\,Rev.\,D} {67}, {063509} {(2003)} [{astro-ph/0209395}].
%
\bibitem{sf-holo}
{X.~Zhang,}  {Int.\,J.\,Mod.\,Phys.\,D} {14}, {063509} {(2005)}
[{astro-ph/0504586}];
%
{M.\,R.~Setare and M.~Jamil,} {Gen.\,Rel.~Gravit.} {43}:{293-303}
{(2011)} [{arXiv:1008.4763}].
%
\bibitem{sf-interact}
{L.\,P.~Chimento, A.\,S.~Jakubi, D.~Pavon and W.~Zimdahl,}
{Phys.\,Rev.\,D}{67}, {083513} {(2003)} [{astro-ph/0303145}];
%
{L.~Zhang, J.~Cui, J.~Zhang, and X.~Zhang,},{Interacting model
of new agegraphic dark energy: Cosmological evolution and
statefinder diagnostic,} {Int.\,J.\,Mod.\,Phys.\,D}{19}, {21} {(2010)} [{arXiv:0911.2838}];
%
{A.~Khodam-Mohammadi and M.~Malekjani,} {Cosmic Behavior,
Statefinder Diagnostic and $w-w^{\prime}$ Analysis for Interacting
NADE model in Non-flat
Universe,} {Astrophys.\,\,Space\,\,Sci.}{331}:{265-273} {(2011)} [{arXiv:1003.0543}].
%
\bibitem{sf-etc}
{Z.\,L.~Yi and T.\,J.~Zhang,} {Statefinder diagnostic for the
modified polytropic Cardassian universe,} {Phys.\,Rev.\,D}{75}, {083515} {(2007)} [{astro-ph/0703630}];
%
{M.~Tong, Y.~Zhang and T.~Xia,} {Statefinder parameters for
quantum effective Yang-Mills condensate dark energy
model,}{Int.\,J.\,Mod.\,Phys.\,D}{18}, {797} {(2009)} [{arXiv:0809.2123}];
%
{H.~Farajollahi and A.~Salehi,} {Attractors, Statefinders and
Observational Measurement for Chameleonic Brans--Dicke
Cosmology,}{JCAP} {1011}:{006} {(2010)} [{arXiv:1010.3589}].
\bibitem{chap1}
A. {Kamenshchik}, U. {Moschella} and V. {Pasquier}, \plb {\bf 511} 265 (2001).

\bibitem{dgp}
G. Dvali, G. Gabadadze and M. Porrati, \plb {\bf 485}, 208 (2000).

\bibitem{faraoni}
V. Faraoni, {Phys.\,Rev.\,D}{62}, {023504} {(2000)} [{arXiv:0002091v2}](2000).

\bibitem{uv} U. Alam and V. Sahni, astro-ph/0209443.
\bibitem{yv}Yuri Shtanov and V. Sahni, JCAP 0311, 014
(2003)[astro-ph/0202346]

\bibitem{STN} We thank A. Starobinsky for bringing this important
result to our notice: A. Starobinsky, Astron. Lett. 7, 36(1981).
\end{thebibliography}
\end{document}